\documentclass[aps,prx,twocolumn,superscriptaddress,final]{revtex4-2}
\usepackage{amsmath}
\usepackage{amsfonts}
\usepackage{amssymb}
\usepackage{bbm}
\usepackage{graphicx}
\usepackage{xcolor}
\usepackage[colorlinks=True,citecolor=blue]{hyperref}
\usepackage{hypcap}
\usepackage{bbold}
\usepackage{braket}
\usepackage{bm}
\usepackage{dsfont}
\usepackage{mathtools}
\usepackage[export]{adjustbox} 
\usepackage{verbatim}

\usepackage{csquotes}
\usepackage[backgroundcolor=white,bordercolor=red,linecolor=red,textcolor=red,textsize=footnotesize]{todonotes}

\begin{document}
\title{Quantum matter is weakly entangled at low energies}
\author{Samuel J. Garratt}
\affiliation{Department of Physics, Princeton University, Princeton, NJ 08544}
\author{Dmitry A. Abanin}
\affiliation{Department of Physics, Princeton University, Princeton, NJ 08544}
\affiliation{\'{E}cole Polytechnique F\'{e}d\'{e}rale de Lausanne (EPFL), 1015 Lausanne, Switzerland}
\affiliation{Google Research, Brandschenkestrasse 150, 8002 Zürich, Switzerland}

\begin{abstract}
We construct upper bounds on entanglement entropies of many-body quantum states that have fixed energy expectation values with respect to local Hamiltonians. Our focus is on entanglement entropies of subsystems that make up approximately half of the full system. The upper bound on the von Neumann entanglement entropy is half the sum of the thermal entropies of two fictitious systems at the same temperature as one another, with an additional area-law contribution in some systems. The effective temperature is chosen such that the sum of the thermal energies of the two fictitious systems matches the constraint on the energy of the state in the original problem; at subextensive energies, this temperature decreases with increasing system size. Our upper bounds on R\'{e}nyi entanglement entropies take an analogous form. These inequalities show that, in systems that are frustration-free (FF) and whose zero-temperature subsystem thermal entropies scale with surface areas (rather than subsystem volumes), ground states are area-law entangled. For physical models of quantum matter, which have well-defined specific heat capacities (and are not necessarily FF), our bounds provide a way to convert this thermodynamic data into constraints on pure-state entanglement at both subextensive and extensive energies. We also show that our general upper bounds on half-system entanglement entropies are optimal, up to subleading corrections, in systems with reflection symmetry and above a subextensive energy threshold. A key application of our results is to constrain the entanglement of quantum states relevant to simulations of zero-temperature response functions, which have energies of order unity. We prove that, in wide varieties of FF systems in two dimensions and higher, the contributions of excitations to von Neumann entanglement entropy are subleading compared with the contribution from the ground state.
\end{abstract}

\maketitle

\section{Introduction}

The study of entanglement in condensed matter systems has provided valuable insights into the complexity of simulations on classical computers. Although generic many-body quantum states have subsystem entanglement entropies scaling with subsystem volumes \cite{page1993average}, and hence cannot be efficiently represented using tensor networks, ground states of local Hamiltonians are typically far less entangled~\cite{holzhey1994geometric,calabrese2004entanglement,refael2004entanglement,fradkin2006entanglement,hastings2007area,gioev2006entanglement,wolf2006violation,metlitski2015entanglement}. 
This is important because many of the coherent collective phenomena observed in experiments can be understood from properties of ground and low-energy excited states, and therefore may be amenable to tensor-network techniques based on low entanglement~\cite{cirac2021matrix}. 

On the other hand, a variety of local Hamiltonians have now been constructed whose ground states are highly entangled \cite{gottesman2010entanglement,irani2010ground,movassagh2016supercritical,zhang2017novel,balasubramanian2023exotic,zhang2023coupled,ippoliti2025infinite,zhang2024lozenge}, violating the area law of subsystem entanglement entropy \cite{hastings2007area} by a power of system size. These results are significant as they appear to rule out the possibility of efficient tensor-network simulations of the ground states of the most general geometrically local Hamiltonians. A basic problem is therefore to identify the physical properties of quantum matter that constrain simulation complexity, in ground states as well as in the low-energy excited states that are important for predicting response functions. 

In this work, we show how thermodynamics constrains the entanglement of many-body pure states in systems with geometrically local interactions. Our central result is a relation between a problem in quantum information theory, the maximization of subsystem von Neumann entanglement entropy subject to a constraint on the global energy expectation value, and a classic problem in statistical physics, in which the total thermal entropy of two independent systems is maximized subject to a constraint on the sum of their energies. This relation leads to an upper bound on subsystem entanglement that is useful for subsystems which make up approximately half of the full system, and that is expressed in terms of subsystem heat capacities. We then construct analogous upper bounds on R\'{e}nyi entanglement entropies.

The idea that thermodynamics should constrain entanglement at finite energy densities is central to the eigenstate thermalization hypothesis \cite{deutsch1991quantum,srednicki1994chaos,deutsch2010thermodynamic}, and more generally to the emergence of quantum statistical mechanics in isolated nonintegrable quantum systems \cite{goldstein2006canonical,popescu2006entanglement}. If a pure state of a large many-body system is locally indistinguishable from a mixed thermal state, then entanglement entropies of small subsystems in the pure state must coincide with thermal entropies. Approximate models for random energy eigenstates have also been used to relate entanglement entropies of large subsystems, constituting up to half of the full system, to thermal entropies \cite{vidmar2017entanglement,lu2019renyi,murthy2019structure,bianchi2019typical}. 

For states with subextensive energy expectation values, Ref.~\cite{brandao2015entanglement} previously showed how von Neumann entropies can be upper bounded by certain thermal entropies. Additionally, for quantum critical systems, Ref.~\cite{swingle2016area} argued that there is a relation between the low-temperature behavior of the thermal entropy and ground-state entanglement. At a technical level, our work goes beyond Refs.~\cite{brandao2015entanglement,swingle2016area} in three important ways. First, in our mappings to thermodynamics problems, we capture the constraint on the energy exactly. Second, our methods apply to both von Neumann \eqref{eq:EEE} and R\'{e}nyi entanglement entropies \eqref{eq:renyibound}. Third, assuming reflection symmetry and also that system boundaries make subleading contributions to thermal entropies, we prove that our upper bounds on half-system (von Neumann and R\'{e}nyi) entanglement entropies are optimal, up to a related subleading correction~\eqref{eq:lowerrenyi}. This statement holds provided the energy above the ground state scales at least as $L^{d-1}$, where $L$ is the system linear extent and $d$ is the spatial dimension. The thermodynamics problems relevant to these results are illustrated in Figs.~\ref{fig:mappingfig1} and \ref{fig:mappingfig}.

The way that we capture the energy constraint is essential for characterizing entanglement in frustration-free (FF) systems, which are defined by the condition that the ground state minimizes the energy with respect to every local contribution to the Hamiltonian. While FF systems are fine-tuned, they are central to our understanding of the quantum computational complexity of preparing many-body quantum ground states. This is because ground states of geometrically local Hamiltonians that are FF up to a boundary condition can encode the outputs of arbitrary quantum circuits \cite{aharonov2008adiabatic,mizel2007simple,breuckmann2014space}. The explicit constructions of highly entangled ground states in Refs.~\cite{irani2010ground,movassagh2016supercritical,zhang2017novel,balasubramanian2023exotic,ippoliti2025infinite} are, moreover, FF up to boundary conditions. Restricting to FF systems has also allowed for significant progress in proving area laws of entanglement in gapped systems in two spatial dimensions~\cite{anshu2022area} and in certain qubit systems with no assumption on the spectral gap \cite{debeaudrap2010solving,chen2011no}. 

Taking the zero-energy limit of our general result, we show that ground-state Schmidt ranks of FF systems are upper bounded by products of ground-state degeneracies of subsystems. If zero-temperature thermal entropies (i.e. logarithms of ground-state degeneracies) of subsystem Hamiltonians are upper bounded by subsystem surface areas, then ground states are area-law entangled. This statement does not depend on the spectral gap, and we note that a closely related upper bound on ground-state entanglement appeared in Ref.~\cite{swingle2016area}. For FF systems, the result is essentially that the third law of thermodynamics implies the area law of ground-state entanglement.

We then apply our upper bounds to study low-energy entanglement in many-body systems with conventional thermodynamic properties, in the sense that their heat capacities are self-averaging. For gapped systems, where the low-temperature specific heat $c(T)$ has an essential singularity at zero temperature $T$, at subextensive energies the half-system entanglement entropy is upper bounded by (i) the product of the total energy, in units of the gap, and (ii) a logarithm of system volume, up to a prefactor~\eqref{eq:gappedbound}. We provide a heuristic justification for this behavior using a quasiparticle picture. For gapless systems whose heat capacities scale as powers of the temperature, the entanglement entropy can be significantly larger than in the gapped case, and we establish the scaling of the upper bound with both energy and system size~\eqref{eq:gaplessbound}. Remarkably, in disordered systems whose heat capacities scale inversely with powers of logarithms of the temperature \cite{fisher1995critical}, and at subextensive energies, our upper bound scales with \emph{volume} up to a multiplicative logarithmic reduction~\eqref{eq:inverselog}. In physical systems which realize such thermodynamic properties, this behavior can be attributed to the emergence of large numbers of non-local low-energy degrees of freedom.

When applied to FF systems, our upper bounds on entanglement separate into contributions from the ground state and from excitations.  This decomposition is extremely useful for understanding the structure of states that only have a finite energy, which arise in particular when ground states are perturbed locally. We prove that, when the low-temperature specific heat capacity is asymptotically smaller than a dimension-dependent power of temperature, $c(T) = o(T^{1/(d-1)})$, the excitation contribution to entanglement is smaller than an area law $L^{d-1}$ ~\eqref{eq:excitationentropy2}. In general it is therefore subleading compared with ground-state entanglement. We anticipate that this application of the bounds will be useful for understanding the complexity of tensor network representation of low-energy states. It is well-known that many FF Hamiltonians have ground states that can be efficiently represented as tensor networks, and our result constrains the entanglement of their excitations.

To support our analytical results we also numerically investigate the entanglement of low-energy eigenstates of non-interacting fermion systems, specifically models involving only nearest-neighbor hopping in one and two spatial dimensions. These models are gapless and have heat capacities that are linear in temperature. Surprisingly, we find that there are families of eigenstates with subextensive energies whose von Neumann entanglement entropies scale with system size according to the same powers predicted by our upper bounds; see Fig.~\ref{fig:metalnumerics}. In one dimension and with only a finite energy above the ground state, there are Slater-determinant eigenstates whose entanglement entropies scale with the square root of system size.

Because energy is conserved under dynamics generated by a fixed Hamiltonian, our work provides limits on the amount of entanglement that can be generated during the evolution of an isolated system. In chaotic systems, evolution starting from a pure initial state is expected to generate a state that is essentially random (although with a fixed energy profile) beyond a time scale set by hydrodynamics \cite{dalessio2016quantum,dymarsky2022bound}. This idea has motivated substantial theoretical work on minimally structured ensembles of many-body quantum states \cite{popescu2006entanglement,goldstein2006canonical,gogolin2016equilibration}, including recent studies \cite{mark2024maximum,teufel2025canonical,mcginley2025scrooge,mok2026nature} of Scrooge ensembles \cite{josza1994lower}. For short-range correlated initial states the total energy is in general a self-averaging quantity, so a baseline expectation for late-time entanglement entropy is provided by the thermal entropy at the corresponding energy. Surprisingly, our work shows that the maximum entanglement of any state (short-range correlated or not) is upper bounded by the thermal entropy associated with that state's energy expectation value, provided this energy is above a subextensive threshold. This last restriction does not apply to FF systems, where the above statement holds for energies arbitrarily close to the ground state.

This paper is organized as follows. In Sec.~\ref{sec:setup} we specify the systems and information-theoretic quantities of interest, and then review basic principles of quantum statistical mechanics, as well as their generalization when von Neumann entropies are replaced by R\'{e}nyi entropies. Our main technical results are in Sec.~\ref{sec:bounds}. There we derive our upper bounds on entanglement entropies for states having fixed energy expectation value. In Sec.~\ref{sec:frustrationfree} we apply these results to ground states of FF systems, and we discuss generic systems at nonzero energies in Sec.~\ref{sec:generic}. In Sec.~\ref{sec:response} we discuss our bounds on entanglement for low-energy states of FF systems, and provide numerical evidence that these results hold more generally. We summarize our results and provide an outlook in Sec.~\ref{sec:discussion}. 

\section{Setup}\label{sec:setup}

The systems that we study can be divided into subsystems $A$, $B$ and $C$, such that $B$ shields $A$ from $C$, i.e. there are no direct interactions between $A$ and $C$, as illustrated in Fig.~\ref{fig:mappingfig1}. The Hilbert space of subregion $A$ is denoted $\mathcal{H}_A$, while its volume is denoted $V_A$ (with analogous notation for other subregions). We also write tensor products of Hilbert spaces as $\mathcal{H}_{AB} \equiv \mathcal{H}_A \otimes \mathcal{H}_B$. In systems with finite-range interactions and total linear extent $L$, we will focus on the case where $V_A,V_C \approx \frac{1}{2}L^d$ and $V_B=O(L^{d-1})$, with $d$ the spatial dimension. In such systems we write the Hamiltonian as
\begin{align}
	H = H_{AB} + H_{BC},
\end{align}
where e.g. $H_{AB}$ is the sum of terms acting in $A$ and $B$ but not in $C$, and so acts on $\mathcal{H}_{AB}$. Note that in general $H_{AB}$ and $H_{BC}$ do not commute. There is some ambiguity in this definition, since for example $H$ will in general involve some terms $H_B$ that only act in $B$. Unless stated otherwise, we choose to split $H_B$ evenly between $H_{AB}$ and $H_{BC}$, i.e. with each having a contribution $\frac{1}{2}H_B$. The ground state energy of $H$ is denoted $E^H_0$, while that of e.g. $H_{AB}$ is denoted $E^{H_{AB}}_0$. 

For a (pure or mixed) state $\rho$ of the system $ABC$, the von Neumann entropy of a subsystem $A$ is defined by
\begin{align}
	S_A(\rho) = -\text{Tr}[ \rho_A \ln \rho_A]
\end{align}
where $\rho_A = \text{Tr}_{BC}\rho$. Due to their importance in tensor-network calculations, it is also useful to consider R\'{e}nyi entropies \cite{schuch2008entropy}, which for index $n \neq 1$ are defined by
\begin{align}
	S_{A,n}(\rho) = (1-n)^{-1} \ln \text{Tr} \big[ \rho_A^{n} \big],
\end{align}
while $S_{A,1}(\rho) \equiv S_A(\rho)$. When the state is pure, $\rho = \ket{\psi}\bra{\psi}$, we will indicate this by instead writing the argument as $\psi$. In this case $S_{A,n}(\psi)$ for all $n$ characterize the entanglement between $A$ and its complement $BC$. 

This work is concerned with the maximum possible values of entanglement entropies $S_A(\psi)$, and more generally $S_{A,n}(\psi)$, subject to a global constraint on the expectation value of the Hamiltonian $E = \braket{\psi|H|\psi}$. We are primarily interested in the leading contributions to the entanglement entropy when $L$ is large and $V_A$ differs from $\frac{1}{2}V$ by no more than $O(L^{d-1})$. These leading contributions will not be sensitive to the fact that $V_A$ is not exactly half of the system volume. 

In the following we will reduce the problem of bounding the entanglement at fixed energy to a thermodynamics problem defined on the overlapping subsystems $AB$ and $BC$. In the rest of this section, we recall basic principles of von Neumann entropy maximization in thermodynamics, along with generalizations to R\'{e}nyi entropies. 

Quantum statistical mechanics is primarily concerned with properties of states $\rho$ that maximize the von Neumann entropy $S_{ABC}(\rho)$ of the entire system subject to a constraint of fixed energy $E=\text{Tr}[\rho H]$. Such Gibbs states take the form
\begin{align}
	\rho^H(T) \propto e^{-H/T}, \label{eq:Gibbs}
\end{align}
with $\text{Tr}\rho^H(T)=1$, where the temperature $T$ is defined as the solution to
\begin{align}
	E = \text{Tr}[H \rho^H(T)]. \label{eq:temperature}
\end{align}
Below it will also be useful to write $E^H(T)$ as the energy expectation value of $\rho^H(T)$, with $E^H(0)=E^H_0$ (the ground state energy). The von Neumann entropy of $\rho^H(T)$ at temperature $T$ is denoted
\begin{align}
	S^H(T) = -\text{Tr}\big[\rho^H(T)\ln \rho^H(T)\big]. \label{eq:thermalentropy1}
\end{align}
In this work, whenever entropies appear with Hamiltonians as superscripts, they are \emph{thermal entropies} for Gibbs states $\rho^H(T)$, i.e. $S^H(T)$ is the entropy of a grand canonical ensemble defined by $H$. It is convenient to write
\begin{align}
	\frac{d S^H}{d E} = \frac{1}{T}, \quad \frac{d E^H}{d T} =  V c^H(T), \label{eq:heatcapacity}
\end{align} 
where the first of these equations is the first law of thermodynamics in the case where the Hamiltonian $H$ is fixed, and the second is the definition of the specific heat capacity $c^H(T)$ of Hamiltonian $H$, here expressed as a function of temperature. 

A similar analysis can be carried out for R\'{e}nyi entanglement entropies. The state $\rho$ of $ABC$ that maximizes the R\'{e}nyi entropy $S_{ABC,n}(\rho)$ subject to $E=\text{Tr}[\rho H]$ is denoted $\rho^H_{n}(T_{n})$. This density matrix is a function of $H$ and of a parameter which we denote by $T_{n}$ (a R\'{e}nyi analog of the temperature) \cite{giudice2021renyi},
\begin{align}
	\rho^H_{n}(T_{n}) \propto \big[ 1 - (n-1)(H-E^H_0)/T_{n}\big]^{1/(n-1)}.\label{eq:RenyiGibbs}
\end{align}
As above, the temperature parameter $T_{n}$ is defined as the solution to $E = \text{Tr}[H \rho^H_{n}(T_{n})]$. Conversely, given $T_{n}$, we have $E^H_{n}(T_{n}) \equiv \text{Tr}[\rho^H_{n}(T_{n})]$. The total R\'{e}nyi entropy of $\rho^H_{n}(T_{n})$ is denoted by
\begin{align}
	S^H_{n}(T_{n}) = (1-n)^{-1} \ln \text{Tr} \big[ \big\{ \rho^H(T_{n})\big\}^{n} \big]. \label{eq:Renyithermalentropy}
\end{align}
We stress that $S^H_{n}(T_{n})$ is \emph{not} the R\'{e}nyi entropy of the Gibbs state $\rho^H(T_{n})$ at temperature $T_{n}$, nor is it the von Neumann entropy of the state $\rho^H_{n}(T_{n})$. The entropy $S^H_{n}(T_{n})$ is the maximum value of $S_{ABC,n}$ subject to $E=\text{Tr}[\rho H]$, and the state which achieves this maximum value is given in Eq.~\eqref{eq:RenyiGibbs}. 

\section{Upper bound on entanglement}\label{sec:bounds}

The amount of entanglement in an isolated quantum system, evolving under a Hamiltonian $H$, will generically increase with time \cite{kim2013ballistic}. Denoting by $E = \braket{\psi|H|\psi}$ the expectation value of the energy, which is fixed, the maximum entanglement of subregion $A$ is
\begin{align}
	\max^{\psi}_{\braket{H}=E} S_A(\psi), \label{eq:maximum}
\end{align}
where the object above `$\max$' is the variable that we optimize over, while the constraint $\braket{H}=\braket{\psi|H|\psi}=E$ is indicated below. We will construct upper bounds on this object, its R\'{e}nyi analogs, and the maximum values of linear combinations of entropies such as 
\begin{align}
	\max^{\psi}_{\braket{H}=E} \big[ S_{A,n}(\psi) + S_{C,n}(\psi)\big], \label{eq:maximum2}
\end{align}
for general $n$. The upper bounds that we derive will be useful when $A$ and $C$ each make up approximately half of the full system, and are otherwise loose. A key motivation for characterizing the quantities in \eqref{eq:maximum} and \eqref{eq:maximum2} in this setting is that they describe the computational resources necessary for tensor-network simulations of the dynamics generated by $H$.

An important technical challenge is to keep track of the constraint on the energy $E$, which can be expressed as a sum of expectation values of $H_{AB}$ and $H_{BC}$. For this reason, our focus will initially be on the object
\begin{align}
	\max^{\rho}_{\text{Tr}[\rho H]=E} \big[ S_{AB,n}(\rho)+S_{BC,n}(\rho)\big]. \label{eq:maximum3}
\end{align}
It is straightforward to see that \eqref{eq:maximum3} is an upper bound on \eqref{eq:maximum2} for all $n$: for pure states, $S_{A,n}(\psi) =S_{BC,n}(\psi)$ and $S_{C,n}(\psi)=S_{AB,n}(\psi)$, and in \eqref{eq:maximum3} we optimize over all states $\rho$ of $ABC$ rather than just pure states. In the specific case of $n=1$, we can use some standard entropic inequalities to upper bound \eqref{eq:maximum}. Using $S_A(\psi) = \frac{1}{2}[S_A(\psi) + S_{BC}(\psi)]$ and the Araki-Lieb inequality $S_A(\psi) \leq S_{AB}(\psi) + S_B(\psi)$ gives
\begin{align}
	S_A(\psi) \leq \frac{1}{2}\big[ S_{AB}(\psi) + S_{BC}(\psi)\big] + \ln \sqrt{|\mathcal{H}_B|}. \label{eq:arakilieb}
\end{align}
Here we have also used $S_B(\psi) \leq \ln |\mathcal{H}_B|$. An inequality of the form \eqref{eq:arakilieb} is only known to hold for $n=1$, and will be useful in this case with the additional constraint that the local Hilbert space dimension is bounded. Our upper bound on \eqref{eq:maximum2}, on the other hand, will hold for all R\'{e}nyi indices $n$. 

We now construct our upper bound on \eqref{eq:maximum3} through a mapping to a problem in thermodynamics. In the interest of grounding our analysis in a familiar problem, it suffices to first focus on $n=1$, where all entropies are standard von Neumann entropies. The first step is to express the problem in \eqref{eq:maximum3}, of maximizing $S_{AB}(\rho) + S_{BC}(\rho)$ subject to $\text{Tr}[\rho H]=E$, as a sequence of two optimization problems. \begin{align}
	&\max^{\rho}_{\text{Tr}[\rho H]=E} \big[S_{AB}(\rho)+S_{BC}(\rho)\big] \label{eq:EAEC} \\
	&= \max^{E_{AB},E_{BC}}_{E_{AB}+E_{BC}=E} \max^{\rho}_{\substack{\text{Tr}[\rho H_{AB}]=E_{AB} \\ \text{Tr}[\rho H_{BC}]=E_{BC}} } \big[S_{AB}(\rho)+S_{BC}(\rho)\big].\notag
\end{align}
The second step, which will allow us to relate von Neumann entropies of subsystems to thermodynamic properties of $H_{AB}$ and $H_{BC}$, is to consider a relaxation of the above problem. Focusing on the inner maximization in Eq.~\eqref{eq:EAEC}, where $E_{AB}$ and $E_{BC}$ are not yet fixed to sum to $E$, we have
\begin{widetext}
\begin{align}
	\max^{\rho}_{\substack{\text{Tr}[\rho_{AB} H_{AB}]=E_{AB} \\ \text{Tr}[\rho_{BC} H_{BC}]=E_{BC}} } \big[S_{AB}(\rho)+S_{BC}(\rho)\big] \leq \max^{\sigma_{AB}}_{\text{Tr}[\sigma_{AB}H_{AB}]=E_{AB}} S_{AB}(\sigma_{AB}) + \max^{\sigma_{BC}}_{\text{Tr}[\sigma_{BC} H_{BC}]=E_{BC}} S_{BC}(\sigma_{BC}). \label{eq:bigineq}
\end{align}
\end{widetext}
On the right-hand side there are two \emph{independent} optimization problems, over $\sigma_{AB}$ and $\sigma_{BC}$ respectively. The point is that we \emph{do not} constrain the density matrices $\sigma_{AB}$ and $\sigma_{BC}$, which are states of $AB$ and $BC$, respectively, to be compatible with any global state of the system. That is, there need not exist some density matrix $\sigma$ of $ABC$ such that $\sigma_{AB} = \text{Tr}_C[\sigma]$ and $\sigma_{BC} = \text{Tr}_A[\sigma]$. Both $\sigma_{AB}$ and $\sigma_{BC}$ are, nevertheless, positive semidefinite matrices with unit trace. The inequality follows from the fact that density matrices of $AB$ and $BC$ that \emph{are} consistent with a global state of $ABC$ are a subset of the density matrices of $AB$ and $BC$ that are \emph{not necessarily} consistent with a global state.

It is important to follow what has happened to our constraint on the energy. On the left-hand side of Eq.~\eqref{eq:bigineq} we have
\begin{align}
	\text{Tr}[\rho_{AB} H_{AB}] = E_{AB} \quad \text{Tr}[\rho_{BC} H_{BC}] = E_{BC}. \notag
\end{align}
On the right-hand side we instead have $\text{Tr}_{AB}[\sigma_{AB} H_{AB}] = E_{AB}$ and $\text{Tr}_{BC}[\sigma_{BC} H_{BC}] = E_{BC}$ with no constraint that $\sigma_{AB}$ and $\sigma_{BC}$ are consistent with one another.

Since the right-hand side of \eqref{eq:bigineq} now involves a maximization of $S_{AB}(\sigma_{AB})$ over $\sigma_{AB}$ subject only to a constraint $\text{Tr}[\sigma_{AB} H_{AB}]=E_{AB}$ that is \emph{local} to subsystem $AB$, the answer is provided to us by statistical mechanics. The optimal $\sigma_{AB}$ is the Gibbs state of the Hamiltonian $H_{AB}$ with energy expectation value $E_{AB}$. Therefore
\begin{align}
	 \max^{\sigma_{AB}}_{\text{Tr}[\sigma_{AB}H_{AB}]=E_{AB}} S_{AB}(\sigma_{AB}) = S^{H_{AB}}(T_{AB}), \label{eq:gibbs2}
\end{align}
where on the right-hand side the temperature $T_{AB}$ is chosen such that
\begin{align}
	E_{AB} = \text{Tr}\big[\rho^{H_{AB}}(T_{AB})H_{AB}\big].
\end{align}
A similar relation holds for the $H_{BC}$ system, and we denote the effective temperature associated with the energy $E_{BC}$ by $T_{BC}$. Note that $S^{H_{AB}}(T_{AB})$ is computed for a system with Hilbert space dimension $\mathcal{H}_{AB}$ (rather than $\mathcal{H}_{ABC}$) and Hamiltonian $H_{AB}$. 

\begin{figure}
\includegraphics[width=0.4\textwidth]{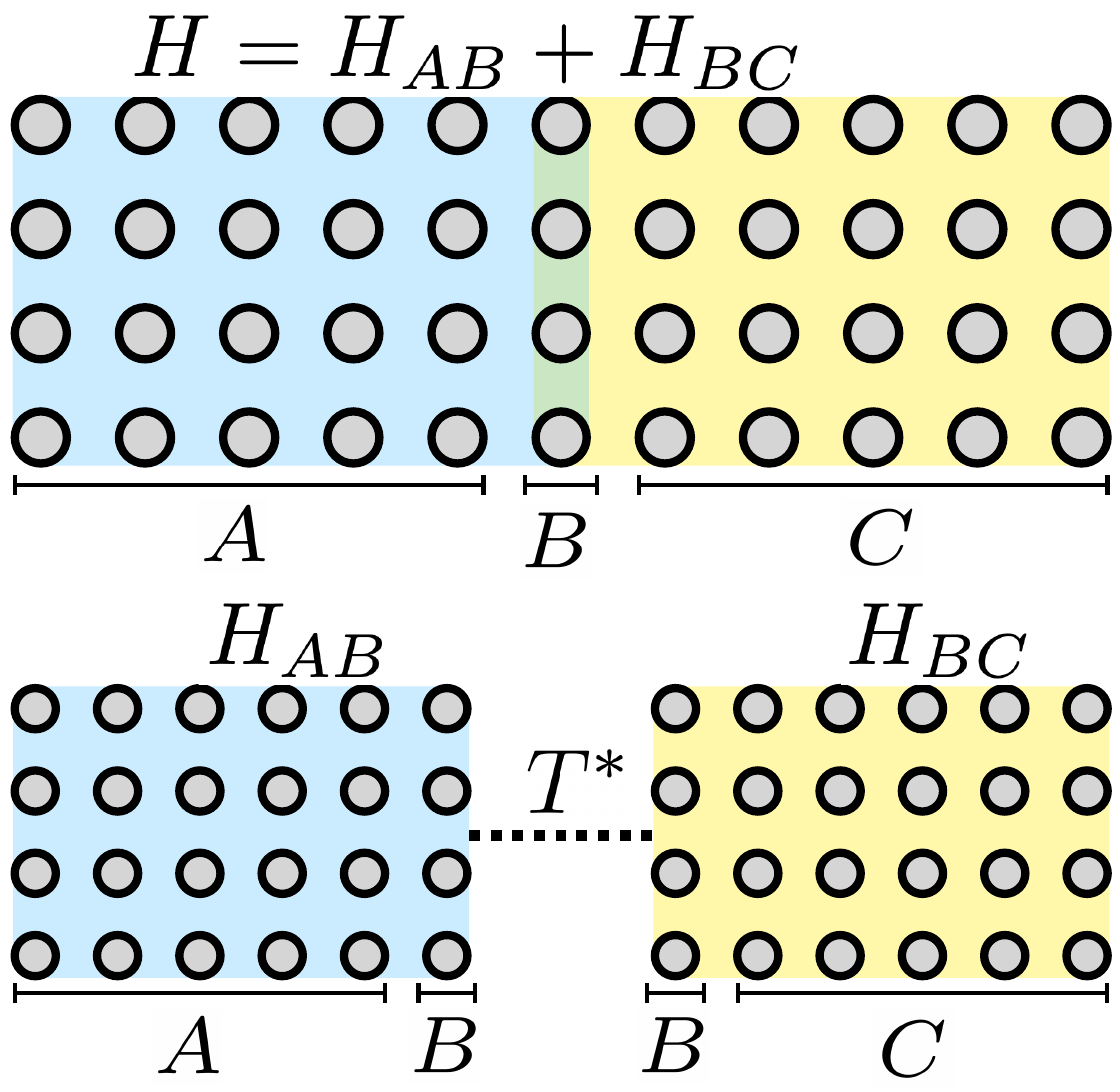}
\caption{Upper bound on entanglement entropy. A quantum system with finite-range interactions can be separated into subregions $A$, $B$ and $C$ such that $A$ and $C$ do not interact. This means that the Hamiltonian $H$ of $ABC$ can be decomposed as $H = H_{AB}+H_{BC}$, where $H_{AB}$ acts only on $A$ and $B$ while $H_{BC}$ acts only on $B$ and $C$. We prove that the entanglement entropy $S_A(\psi)$ of a pure state $\ket{\psi}$ with energy expectation value $E = \braket{\psi|H|\psi}$ is upper bounded by the sum of the thermal entropies of two fictitious systems having Hamiltonians $H_{AB}$ and $H_{BC}$ acting on \emph{distinct} copies of the $B$ degrees of freedom, as well as an area-law contribution. The fictitious $H_{AB}$ and $H_{BC}$ systems are in thermal equilibrium with one another at a temperature $T^*$, with $T^*$ determined by the condition that $E_{AB}+E_{BC}=E$, where $E_{AB}$ ($E_{BC}$) is the energy expectation value of the thermal state of $H_{AB}$ ($H_{BC}$) at $T^*$. Analogous upper bounds apply for R\'{e}nyi entanglement entropies.}
\label{fig:mappingfig1}
\end{figure}

Inserting \eqref{eq:gibbs2} into \eqref{eq:bigineq} we find an upper bound on $S_{AB}(\rho)+S_{BC}(\rho)$ subject to the constraints $\text{Tr}[\rho H_{AB}]=E_{AB}$ and $\text{Tr}[\rho H_{BC}]=E_{BC}$. Then, inserting \eqref{eq:bigineq} into the outer maximization on the right-hand side of \eqref{eq:EAEC}, we find an upper bound on $S_{AB}(\rho)+S_{BC}(\rho)$ subject to the single \emph{global} constraint $E = \text{Tr}[\rho H]$. This upper bound is
\begin{align}
	&\max^{\rho}_{\text{Tr}[\rho H]=E} \big[S_{AB}(\rho)+S_{BC}(\rho)\big] \notag \\
	&\leq \max^{E_{AB}}_{E_{AB}+E_{BC}=E} \big[S^{H_{AB}}(T_{AB}) + S^{H_{BC}}(T_{BC})\big]. \label{eq:summary1}
\end{align}
The final step is to maximize the sum of the thermal entropies $S^{H_{AB}}(T_{AB}) + S^{H_{BC}}(T_{BC})$ subject to the constraint $E=E_{AB}+E_{BC}$. The solution to such problems is well known: we should set the temperatures, defined relative to $H_{AB}$ and $H_{BC}$, to be equal to one another. We denote the solution by $T^* \equiv T_{AB} = T_{BC}$, where
\begin{align}
	E = \text{Tr}\big[ \rho^{H_{AB}}(T^*) H_{AB}\big] +  \text{Tr}\big[ \rho^{H_{BC}}(T^*)  H_{BC}\big].  \label{eq:equilibriumcondition}
\end{align}
Equation~\eqref{eq:equilibriumcondition} appears to describe two systems in thermal contact, but we cannot view the systems having Hamiltonians $H_{AB}$ and $H_{BC}$ as subsystems of the overall system $ABC$: each has its own `copy' of subsystem $B$, as illustrated in Fig.~\ref{fig:mappingfig1}. We then find\\
\textbf{Theorem 1}
\begin{align}
	\max^{\rho}_{\text{Tr}[\rho H]=E} &\big[S_{AB}(\rho)+S_{BC}(\rho)\big]\notag \\ &\leq S^{H_{AB}}(T^*)  + S^{H_{BC}}(T^*),  \label{eq:EEE}
\end{align}
\emph{where} $T^*$ \emph{is the solution to} $E = E^{H_{AB}}(T^*)+E^{H_{BC}}(T^*)$\emph{, and on the left-hand side the optimization is performed over all mixed states $\rho$ of $ABC$.}

The above analysis is modified only slightly for general R\'{e}nyi indices $n$, departing from the result for von Neumann entropies only after \eqref{eq:summary1}. While maximizing a sum of entropies subject to a constraint on the total energy led us to set $T_{AB}=T_{BC}$ in the case of von Neumann entropies, this is no longer the case for R\'{e}nyi entropies with $n \neq 1$. For this reason our general result does not simplify beyond\\ \textbf{Theorem 2}
\begin{align}
	&\max^{\rho}_{\text{Tr}[\rho H]=E} \big[S_{AB,n}(\rho)+S_{BC,n}(\rho)\big] \notag \\
	&\leq \max^{E_{AB}}_{E_{AB}+E_{BC}=E} \big[S^{H_{AB}}_{n}(T_{AB,n}) + S^{H_{BC}}_{n}(T_{BC,n})\big], \label{eq:renyibound}
\end{align}
\emph{where $T_{AB,n}$ and $T_{BC,n}$ are R\'{e}nyi analogs of temperature chosen such that $E=E^{H_{AB}}_{n}(T_{AB,n})+E^{H_{BC}}_{n}(T_{BC,n})$.}

Returning to the case of von Neumann entropies, we now convert \eqref{eq:EEE} into an upper bound on the entanglement entropy of a subsystem $A$ of a pure state $\ket{\psi}$ of $ABC$. Combining \eqref{eq:arakilieb} and \eqref{eq:EEE}, we find\\
\textbf{Corollary 1}
\begin{align}
	\max^{\psi}_{\braket{H}=E} S_A(\psi) \leq \frac{1}{2}\Big[ &S^{H_{AB}}( T^*) + S^{H_{BC}}( T^*)\Big]  \label{eq:psibound1}  \\&+ \ln \sqrt{|\mathcal{H}|_B}\notag .
\end{align}
We also have \\
\textbf{Corollary 2}
\begin{align}
	\max^{\psi}_{\braket{H}=E} \big[ S_A(\psi) + S_C(\psi)\big] \leq &S^{H_{AB}}( T^*) + S^{H_{BC}}( T^*) \label{eq:psibound2},
\end{align}
which has the advantage of being applicable to systems with both bounded and unbounded local Hilbert space dimensions. The result in \eqref{eq:psibound2} for von Neumann entropies has a simple analog for R\'{e}nyi entropies, in which one replaces the left-hand side by $S_{A,n}(\psi) + S_{C,n}(\psi)$ for general $n$, and the upper bound by the right-hand side of \eqref{eq:renyibound}.

Finally we note that the above bounds simplify in physical settings in which both the states and Hamiltonians are symmetric under reflection about $B$, such that $A$ and $C$ are exchanged. With this additional reflection symmetry imposed we find\\
\textbf{Corollary 3}
\begin{align}
	\max^{\psi}_{\braket{H}=E} S_{A,n}(\psi) \leq  S^{H_{AB}}_{n}(T^*_{n}),\label{eq:reflection}
\end{align}
for all $n$ and for both bounded and unbounded local Hilbert space dimensions. For reflection symmetric systems we can also construct a lower bound on the left-hand side of \eqref{eq:reflection} in terms of the solution to a different fictitious thermodynamics problem, which differs from the one on the right-hand side of \eqref{eq:reflection} only in its boundary conditions, and we discuss this in the next Section. In systems where the boundary makes only a subleading contribution to the thermal entropy at large $L$, \eqref{eq:reflection} is therefore optimal up to these boundary corrections. 

In this Section we have constructed upper bounds on entanglement in terms of the solution to a thermodynamics problem. The right-hand side of \eqref{eq:EEE} is the total thermal entropy of two fictitious systems, with Hamiltonians $H_{AB}$ and $H_{BC}$, respectively, both at a temperature $T^*$ such that the sum of their energies is $E$. The physical interpretation of \eqref{eq:psibound1} and \eqref{eq:psibound2} is simply that, for pure states of $ABC$, we cannot increase the entanglement entropy of $A$ without increasing the entanglement of $BC$.

\section{Lower bound on entanglement}\label{sec:lowerbound}

Here we derive simple lower bounds on the maximum possible values of $S_{A,n}(\psi)$ for states with $\braket{\psi|H|\psi}=E$, for general R\'{e}nyi indices $n$. These bounds apply to systems with reflection symmetry and in a geometry similar to Fig.~\ref{fig:mappingfig1}, so that e.g. $H_{AB}$ and $H_{BC}$ are equal to one another (although they act on different spaces). We also require $E = E^H_0+ \Omega(L^{d-1})$. Crucially, the lower bound on the entanglement of $A$ is the thermal entropy of a system that differs from $H_{AB}$ only in its boundary conditions. Typically a change in boundary conditions leads only to a subleading change in the entropy. The results of this Section then imply that the upper bounds derived in the previous Section are optimal up to corrections that are subleading at large $L$. 

\begin{figure}
\includegraphics[width=0.4\textwidth]{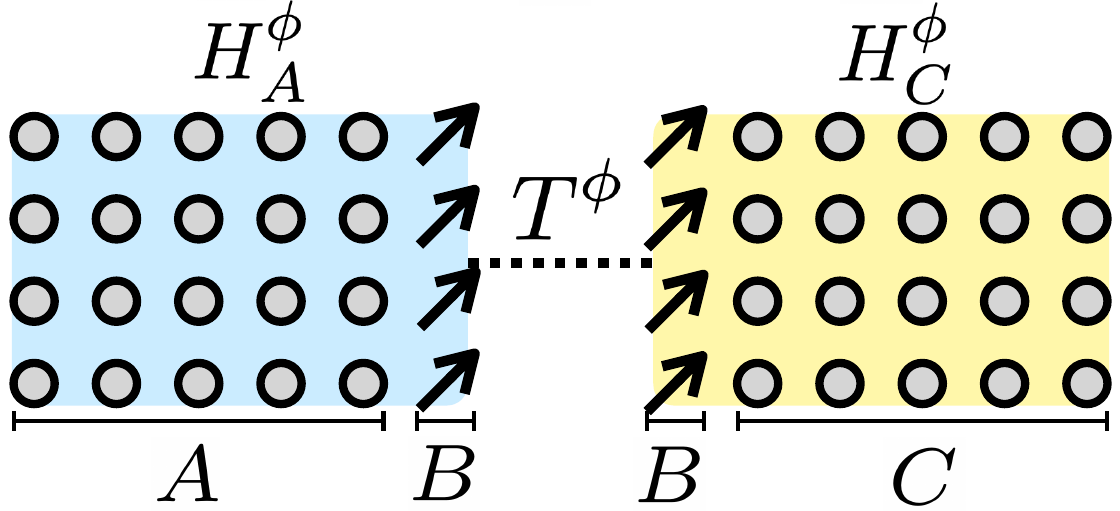}
\caption{Lower bounds on maximum entanglement entropy in reflection-symmetric systems. The Hamiltonian $H$ of the system $ABC$ can be decomposed as $H = H_{AB}+H_{BC}$, where $H_{AB}$ acts only on $A$ and $B$ while $H_{BC}$ acts only on $B$ and $C$ (see Fig.~\ref{fig:mappingfig1}). Here $A$ and $C$ are related by a reflection symmetry represented by the unitary operator $R$, which acts as $R H_{AB} R^{-1} = H_{BC}$. The lower bound on the maximum possible (von Neumann or R\'{e}nyi) entanglement entropy $S_{A,n}(\psi)$ of subsystem $A$ in a state $\ket{\psi}$ with fixed $E =\braket{\psi|H\psi}$ is given by half the sum of thermal (Gibbs or R\'{e}nyi) entropies of two systems with the structure shown in the lower panel. There the $B$ degrees of freedom are fixed to an arbitrary state $\ket{\phi}$ and the effective Hamiltonians for $A$ and for $C$ are given in Eq.~\eqref{eq:Hphi}. The effective temperature $T^{\phi}_{n}$ of the systems in the lower panel is fixed by the condition in Eq.~\eqref{eq:Tphi}.}
\label{fig:mappingfig}
\end{figure}

Recall that $H_{AB}$ ($H_{BC}$) does not act on degrees of freedom in $C$ ($A$), and that the subregion $B$ can be chosen somewhat arbitrarily as long as it has volume $O(L^{d-1})$. The lower bound on the maximum entanglement entropy at fixed energy \eqref{eq:maximum} follows from a specific choice of a pure state of $ABC$,
\begin{align}
	\ket{\psi} \propto \Big(e^{-H^{\phi}_A/4T^{\phi}} \otimes e^{-H^{\phi}_C/4T^{\phi}} \ket{\text{EPR}}_{AC}\Big) \otimes \ket{\phi}_B, \label{eq:psiphi}
\end{align}
where $\ket{\text{EPR}}_{AC}$ is a maximally entangled state shared between $A$ and $C$, and 
\begin{align}
	H^{\phi}_A = \braket{\phi|H_{AB}|\phi}, \quad H^{\phi}_C = \braket{\phi|H_{BC}|\phi}. \label{eq:Hphi}
\end{align} 
The Hamiltonians $H^{\phi}_A$  and $H^{\phi}_C$ act only on $A$ and on $C$ degrees of freedom, respectively, and are constructed by imposing that the $B$ degrees of freedom are in a state $\ket{\phi}$. For simplicity, we can choose $\ket{\phi}$ to be a product state over the $B$ degrees of freedom, in which case the interactions between e.g. $A$ and $B$ degrees of freedom in $H_{AB}$ are replaced by interactions between $A$ and `boundary fields' in $H_A^{\phi}$. In this case $H^{\phi}_A$ is a Hamiltonian for $A$ in which quantum fluctuations of $B$ have been eliminated. It can be verified that
\begin{align}
	\braket{\psi|H_{AB}|\psi} = E^{H^{\phi}_A}(T^{\phi}),
\end{align}
where $E^{H^{\phi}_A}(T^{\phi})$ is the energy of a state that is thermal with respect to $H^{\phi}_A$ and at temperature $T^{\phi}$. To construct a lower bound that applies for a total system energy of $E$, we must therefore choose $T^{\phi}$ such that
\begin{align}
	E = E^{H^{\phi}_A}(T^{\phi}) + E^{H^{\phi}_C}(T^{\phi}), \label{eq:Tphi}
\end{align}
and we note that the two terms on the right-hand side of Eq.~\eqref{eq:Tphi} are equal since the system is reflection symmetric about $B$. A solution to Eq.~\eqref{eq:Tphi} will generically only exist for $E-E^H_0 = \Omega(L^{d-1})$. In this regime, for the specific choice of state in \eqref{eq:psiphi} we have $S_A(\psi) = S^{H_A^{\phi}}(T^{\phi})$, so our lower bound is
\begin{align}
	\max^{\psi}_{\braket{H}=E} S_A(\psi) \geq S^{H_A^{\phi}}(T^{\phi}).\label{eq:lower}
\end{align}
Under these symmetry restrictions the upper bound from \eqref{eq:psibound1} is simply $S^{H_{AB}}(T^*)$, with $T^*$ chosen such that $E = 2E^{H_{AB}}(T^*)$. The setup relevant to the lower bound in \eqref{eq:lower} is illustrated in Fig.~\ref{fig:mappingfig}. We therefore have upper and lower bounds on the entanglement entropy of $A$ that are expressed in terms of thermal entropies with respect to Hamiltonians $H_A^{\phi}$ and $H_{AB}$ which only differ where they involve $B$ degrees of freedom (i.e. in the boundary region). For large system sizes and $E=\Omega(L^{d-1})$ the temperatures $T^{\phi}$ and $T^*$ will approach one another, and the thermal entropies $S^{H_A^{\phi}}(T^{\phi})$ and $S^{H_{AB}}(T^{\phi})$ will differ only by a contribution that is smaller than the leading term by a factor $\sim V_B/V_A = O(L^{-1})$. Unless the Hamiltonian has highly unconventional thermodynamic properties (for example if its thermal entropy is dominated by a surface term), the upper bound in \eqref{eq:psibound1} is optimal up to a subleading correction. 

This construction immediately generalizes to R\'{e}nyi entropies. For R\'{e}nyi index $n \neq 1$, define
\begin{align}
    &\ket{\psi} \propto \Big( \big[\rho^{H^{\phi}_A}(T^{\phi}_{n})\big]^{1/4} \otimes \big[\rho^{H^{\phi}_C}(T^{\phi}_{n})\big]^{1/4} \ket{\text{EPR}}_{AC}\Big) \otimes \ket{\phi}_{B}\notag \\
	&\rho^{H^{\phi}_A}(T^{\phi}_{n}) \propto \big[ 1 - (n-1)(H^{\phi}_A-E^{H^{\phi}_A}_0)/T^{\phi}_{n}\big]^{1/(n-1)},
\end{align}
where the second line is Eq.~\eqref{eq:RenyiGibbs} for the Hamiltonian $H^{\phi}_A$, and there is an analogous expression for $\rho^{H^{\phi}_C}(T^{\phi}_{n})$. As above, we require $H^{\phi}_A$ and $H^{\phi}_C$ to be isospectral. The R\'{e}nyi temperature is then fixed by the condition
\begin{align}
    E = E^{H^{\phi}_A}_{n}(T^{\phi}_{n}) + E^{H^{\phi}_C}_{n}(T^{\phi}_{n}),
\end{align}
which is analogous to Eq.~\eqref{eq:Tphi}. When a solution to this equation exists, i.e. for $E-E^H_0 = \Omega(L^{d-1})$, we have
\begin{align}
	\max^{\psi}_{\braket{\psi|H|\psi}=E} S_{A,n}(\psi) \geq S_{n}^{H_A^{\phi}}(T^{\phi}_{n}), \label{eq:lowerrenyi}
\end{align}
where the right-hand side is the R\'{e}nyi entropy of the density matrix $\rho^{H^{\phi}_A}(T^{\phi}_{n})$ of $A$, defined in Eq.~\eqref{eq:Renyithermalentropy}.

In this section we have shown that the maximum possible von Neumann and R\'{e}nyi entropies at fixed energy $E-E^H_0 = \Omega(L^{d-1})$ can be \emph{lower bounded} by thermal entropies in fictitious thermodynamics problems that are closely related to ones appearing in the upper bounds in Sec.~\ref{sec:bounds}. The thermodynamics problems associated with the lower bounds differ from those associated with the upper bounds only in and adjacent to the subregion $B$, which makes up a vanishing fraction of the system volume. Provided this boundary region does not dominate thermodynamic properties, our upper bounds on half-system entanglement entropies are optimal up to subleading additive corrections. 

\section{Frustration-free ground states}\label{sec:frustrationfree}

Before applying the inequality \eqref{eq:psibound1} to generic geometrically local systems, in this Section we use it to derive simple upper bounds on the entanglement entropies of ground states of FF systems. FF Hamiltonians have the form $H = \sum_j h_j$, where each geometrically local term $h_j$ annihilates the ground state $h_j \ket{\psi^H_0}=0$. The ground state energy is then $E^H_0=0$. 

There are several motivations for studying FF systems in this setting. First, although the ground states of generic local Hamiltonians appear to be quite weakly entangled, a wide variety of FF models have been constructed as counterexamples to this expectation. Notably, Refs.~\cite{movassagh2016supercritical,zhang2017novel} constructed gapless FF Hamiltonians for one-dimensional qudit systems whose unique ground states $\ket{\psi^H_0}$ have half-system entanglement entropies scaling with system length $L$ as $S_A(\psi^H_0) \sim L^{1/2}$. This is to be contrasted with the behavior $S_A(\psi^H_0) \sim \ln L$ typically expected in ground states of gapless Hamiltonians \cite{calabrese2004entanglement}. Two-dimensional generalizations of this construction have been shown  \cite{zhang2023coupled,zhang2024lozenge} and argued \cite{balasubramanian2023exotic} to have unique volume-law entangled ground states $S_A(\psi^H_0) \sim L^2$. 

The second motivation is related to the conjecture that ground states of geometrically local and gapped local Hamiltonians have entanglement entropies that scale with the areas of subsystem boundaries. Ref.~\cite{hastings2007area} proved this conjecture for translation-invariant and gapped spin systems in one spatial dimension, but beyond one spatial dimension the only current proofs of area-law entanglement apply to FF systems \cite{beaudrap2010ground,chen2011no,anshu2022area}. In particular, Refs.~\cite{beaudrap2010ground,chen2011no} proved area-law entanglement for families of qubit systems with nearest-neighbor interactions, without any assumption on the spectral gap. The proof of Ref.~\cite{anshu2022area} applies to a broader class of Hamiltonians but operates under the assumption that the spectral gap of any contiguous subset of terms in the Hamiltonian has a finite gap. 

In Sec.~\ref{sec:FFbound} we construct a gap-independent upper bound on the Schmidt ranks of ground states of FF systems. This bound holds in general spatial dimensions. In Sec.~\ref{sec:AKLT} and \ref{sec:Motzkin} we apply this result to paradigmatic FF spin chains with entangled ground states, and in Sec.~\ref{sec:Glauber} we apply it to a family of FF Hamiltonians (in general dimensions) whose ground states can be exactly represented as projected entangled pair states (PEPS). In all of these cases our upper bound is saturated (up to corrections that are subleading at large $L$). 

\subsection{Upper bound on ground-state entanglement}\label{sec:FFbound}

Our results on the entanglement of the ground states of FF systems follows immediately from our upper bound on the sum of R\'{e}nyi entropies in \eqref{eq:renyibound}. First recall that, in the decomposition $H = H_{AB} + H_{BC}$ of the Hamiltonian, we previously had some freedom as to how to divide the terms $H_B$, which act only in $B$, between $H_{AB}$ and $H_{BC}$. Here it is important that each of $H_{AB}$ and $H_{BC}$ has a positive contribution from $H_B$, i.e. 
\begin{align}
	H_{AB} = \cdots + \lambda H_B, \quad H_{BC} = \cdots + (1-\lambda)H_B,
\end{align}
with $0 < \lambda < 1$, where in $H_{AB}$ ($H_{BC}$) the ellipsis denotes terms $h_j$ supported partially on $A$ ($C$). For such a split, each of $H_{AB}$ and $H_{BC}$ are also frustration-free with ground state energies $E^{H_{AB}}_0=E^{H_{BC}}_0=0$. The ground state degeneracies $D_{AB}$ and $D_{BC}$ of each of $H_{AB}$ and $H_{BC}$ are, moreover, independent of how we split up $H_B$ provided $0 < \lambda < 1$. 

Now we set $E=0$ in \eqref{eq:psibound2}; the analysis is the same for the R\'{e}nyi analog of this inequality discussed below \eqref{eq:psibound2}. To achieve $E=0$ on the right-hand side of \eqref{eq:psibound2} and its R\'{e}nyi generalization \eqref{eq:renyibound}, the only possibility is that the fictitious thermodynamics problems appearing on the right-hand sides of the inequalities are at zero temperature, $T^*_{AB,n}=T^*_{BC,n}=0$. The thermal entropies are then fixed by the ground state degeneracies, $S^{H_{AB}}_{n}(0)=\ln D_{AB}$ and $S^{H_{BC}}_{n}(0)=\ln D_{BC}$ for all $n$. This leads to\\
\begin{align}
	S_{A,n}(\psi^H_0)+S_{C,n}(\psi^H_0) \leq \ln \big[ D_{AB}D_{BC} \big] , \label{eq:FFbound}
\end{align}
which applies for any ground state $\ket{\psi^H_0}$ of a frustration-free Hamiltonian $H = H_{AB}+H_{BC}$. 

This result is very intuitive, and we can arrive at a closely related upper bound following Ref.~\cite{swingle2016area}. In a FF system, any ground state of $H$ is also a ground state of $H_A$. Therefore, when constructing ground states of $H$, we can only make use of states within a $D_A$-dimensional subspace of $\mathcal{H}_{A}$, where $D_A$ is the ground-state degeneracy of $H_A$ within the space $\mathcal{H}_A$. This restriction means that $S_{A,n}(\psi^H_0)$ is upper bounded by the minimum of $\ln D_A$ and $\ln D_{BC}$~\cite{swingle2016area}. The specific form of our result in \eqref{eq:FFbound} is useful because it can be extended to nonzero energies; it is the $E \to E^H_0$ limit of the upper bounds that we will discuss in Sec.~\ref{sec:generic}.

The result \eqref{eq:FFbound} and the discussion below apply for all $n$ including $n=0$, i.e. they constrain the Schmidt rank of the ground state $\ket{\psi^H_0}$ across a bipartition into subregions $A$ and $BC$. Additionally, this relation between entanglement and ground-state degeneracy does not depend at all on the spectral gap. It is important to note that the ground-state degeneracies of quantum systems can depend sensitively on boundary conditions, and \eqref{eq:FFbound} shows that for $H$ with a unique ground state such sensitivity is essential if that ground state is also entangled. We illustrate this point using concrete examples below. 

A key feature of \eqref{eq:FFbound} is that, when we construct the fictitious subsystems with Hamiltonians $H_{AB}$ and $H_{BC}$, there will be generically be local terms `missing' at the boundary; we will see an example in Sec.~\ref{sec:Glauber}. These missing terms can cause $D_{AB}$ and $D_{BC}$ to grow exponentially with $L^{d-1}$, the volume that is within the interaction range of subsystem $B$. In particular, if $\ln D_{AB},\ln D_{BC} = O(L^{d-1})$ then from \eqref{eq:FFbound},
\begin{align}
	S_{A,n}(\psi^H_0) = O(L^{d-1}), \label{eq:FFarea}
\end{align}
for all $n$, which is area-law entanglement. Quite generally, in physical systems one expects that the ground-state degeneracy should be controlled by the total surface area of the system, rather than by the system volume. In FF systems, this condition on the zero-temperature thermal entropy \emph{implies} the area law of entanglement entropy.  

This result becomes particularly strong for FF Hamiltonians whose ground states are unique in geometries with open boundary conditions. To be concrete, we consider cubic lattices in general dimensions, where qudits are at the vertices, and where all interactions are between qudits within the same unit cell. This includes, for example, interactions involving multiple qudits around a plaquette on a square lattice, as was studied in Ref.~\cite{anshu2022area}. If the ground state of such a FF Hamiltonian is \emph{unique} in geometries with open boundary conditions, then
\begin{align}
	S_{A,n}(\psi^H_0) = 0,
\end{align}
meaning that the ground states are product states. In the following, we discuss the application of \eqref{eq:FFbound} to two kinds of quantum spin chains with entangled ground states and to a family of two-dimensional quantum Ising models.

\subsection{AKLT chain}\label{sec:AKLT}

As an example of a model with an area-law entangled ground state, let us consider a paradigmatic model for symmetry-protected topological (SPT) order. The Affleck-Kennedy-Lieb-Tasaki (AKLT) model \cite{affleck1988valence} describes spin-one degrees of freedom (qutrits) with nearest-neighbor interactions in one spatial dimension, and with \emph{periodic} boundary conditions its ground state is unique. The full Hamiltonian is
\begin{align}
	H = \sum_{j=0}^{L-1} \Pi^{\text{AKLT}}_{j,j+1}, \label{eq:AKLT}
\end{align}
where $\Pi_{j,j+1}$ is a projector into the spin-two subspace of two adjacent spin-one degrees of freedom, here indexed by $j$ and $j+1$, with $\Pi_{L-1,L} \equiv \Pi_{L-1,0}$. Let us choose $A$ to consist of sites $j=0, \ldots, \lfloor L/2 \rfloor$, $B$ to consist of the two sites $j=\lfloor L/2 \rfloor+1$ and $j=L-1$ on opposite sides of the system, and $C$ to consist of all other sites. The Hamiltonian $H_{AB}$ is then the sum of $\Pi^{\text{AKLT}}_{j,j+1}$ with $j \leq \lfloor L/2 \rfloor$.

In this setup, each of $H_{AB}$ and $H_{BC}$ are AKLT models with \emph{open} boundary conditions, which have ground degeneracies $D_{AB}=D_{BC}=4$. This degeneracy is typically understood as arising from degrees of freedom on the boundary. Using \eqref{eq:FFbound} we then find the upper bound $S_A(\psi^H_0)+S_C(\psi^H_0) \leq 4 \ln 2$. However, for even $L$ the AKLT model (and its ground state with PBCs) is invariant under exchange of $A$ and $C$, so that
\begin{align}
	S_{A,n}(\psi^H_0) \leq 2 \ln 2,
\end{align}
which is saturated at large $L$ \cite{hirano2007entanglement}. This example is interesting in part because the AKLT state is just one representative of a SPT phase. Although other Hamiltonians with ground states in this phase are not necessarily FF, there is still a relation between entanglement and (approximate) ground-state degeneracy \cite{pollmann2010entanglement}. 

\subsection{Motzkin chain}\label{sec:Motzkin}

As a second example we discuss the Motzkin chain introduced in Ref.~\cite{bravyi2012criticality}. Here we focus on the `colorless' version of the model, whose ground state has half-system entanglement entropy growing logarithmically with system size. `Colorful' versions of the model~\cite{movassagh2016supercritical,zhang2017novel} have entanglement entropies that instead grow as powers of system size. Our analysis of the colorless version will provide insights into the origin of entanglement in both classes of models. 

The colorless Motzkin chain consists of qutrits (spin one degrees of freedom) with nearest-neighbor interactions along a one-dimensional lattice. With open boundary conditions the ground state is unique and has half-system entanglement entropy $S_A(\psi^H_0) = \frac{1}{2}\ln L + O(1)$. Crucially, the fact that the ground state is unique depends on the introduction of boundary terms on qutrits at the ends of the chain. Writing our orthonormal basis for states of individual qudits as $\ket{-1}$, $\ket{0}$, $\ket{1}$, the Hamiltonian is 
\begin{align}
	H = \sum_{j=0}^{L-2} \Pi^{\text{Motz.}}_{j,j+1} + \ket{-1}\bra{-1}_0 + \ket{1}\bra{1}_{L-1},
\end{align}
where we index qutrits by $j=0,1,\ldots, L-1$, and $\Pi^{\text{Motz.}}_{j,j+1}$ is a rank-three projector. This projector takes the form $\Pi^{\text{Motz.}} = \sum_{a=1}^3 \ket{\chi_a}\bra{\chi_a}$ with
\begin{align}
	\ket{\chi_1} &\propto \ket{0,0} - \ket{1,-1},\notag\\
	\ket{\chi_2} &\propto \ket{1,0} - \ket{0,1},\\
	\ket{\chi_3} &\propto \ket{-1,0} - \ket{0,-1}.\notag
\end{align}
In the absence of either one (or both) of the boundary terms $\ket{-1}\bra{-1}_0$ and $\ket{1}\bra{1}_{L-1}$, the Hamiltonian $H$ would have a ground state degeneracy of order $L$~\cite{bravyi2012criticality}. Introducing `color' to the Motzkin chain~\cite{movassagh2016supercritical,zhang2017novel} enhances this degeneracy to be exponential in $L$. From \eqref{eq:FFbound} we can now see that such degeneracies are inextricably linked to the large entanglement in the unique ground states of Motzkin chains. 

Let us choose $A$ to consist of the first $\lfloor L/2\rfloor-1$ sites, $B$ to be the site adjacent to $A$, and $C$ to be the remaining sites. Then, $H_{AB}$ and $H_{BC}$ each take the forms of Motzkin chains with only one boundary term, i.e. $\ket{-1}\bra{-1}_0$ in $H_{AB}$ and $\ket{1}\bra{1}_{L-1}$ in $H_{BC}$. As a consequence, $D_{AB}$ and $D_{BC}$ are each of order $L$ in the colorless model. The degeneracy that arises from the absence of boundary terms gives $S_{A,n}(\psi^H_0)+S_{C,n}(\psi^H_0) < 2\ln L + O(1)$. Using the symmetry under reflection about $B$, we then have
\begin{align}
	S_{A,n}(\psi^H_0) < \ln L + O(1),
\end{align}
for the unique ground state of the colorless Motzkin chain. For $n=0$ this upper bound is optimal up to a correction that is subleading at large $L$.

The above addresses a question, raised in previous work on FF systems~\cite{bravyi2012criticality}, of whether highly entangled and unique ground states can arise without requiring unusual boundary conditions. Our analysis shows that this is not possible in FF systems. A half-system entanglement entropy that grows rapidly with system size requires that $H_{AB}$ and $H_{BC}$, which necessarily differ from $H$ in their boundary conditions, have ground-state degeneracies that grow with subsystem size. 

\subsection{Rohksar-Kivelson models}\label{sec:Glauber}

Here we illustrate how area-law entanglement arises in the ground states of generic FF models. To highlight the fact that this entanglement is essentially independent of the existence of a spectral gap, we will discuss a model whose spectral gap undergoes a transition as a parameter is tuned, but whose bipartite entanglement entropy follows an area law for all (nonzero) values of this parameter. For a system of qubits on a cubic lattice with open boundary conditions, consider 
\begin{align}
	H = \sum_j\big( e^{-\beta \sum_{k|j} Z_j Z_k} - X_j\big)\label{eq:GlauberH}
\end{align}
where $\sum_{k|j}$ denotes a sum over the $2d$ sites $k$ of the lattice that are adjacent to site $j$, and each of $X_j$ and $Z_j$ are standard Pauli operators. The Hamiltonian is related by a similarity transformation to the generator of classical stochastic Glauber dynamics for a ferromagnetic Ising model on a square lattice \cite{verstraete2006criticality}, and so belongs to a broader family of models that includes the Rohksar-Kivelson quantum dimer Hamiltonians \cite{rokhsar1988superconductivity}. In a finite system the ground state is unique and takes the form
\begin{align}
	\ket{\psi^H_0} \propto \sum_{ \{ z\}} e^{(\beta/4)\sum_j \sum_{k | j} z_j z_k} \ket{\{ z\}},
\end{align}
where $Z_j \ket{\{ z\}} = z_j \ket{\{ z\}}$ for all $j$, with $z_j = \pm 1$. For $d \geq 2$ and $\beta$ larger than a $d$-dependent critical value $\beta_c$ the spectral gap separating the two quasi-degenerate ground states is exponentially small in $L$, and excitations above these states are gapless \cite{huse1987dynamics,sahay2025enforced}. Here it will suffice to consider large finite $L$ (rather than strictly infinite $L$), and this will allow us to make concrete statements about the entanglement of the unique ground state $\ket{\psi^H_0}$. This state can be exactly represented as PEPS with bond dimension two \cite{verstraete2006criticality}. 

\begin{figure}
\centering
\includegraphics[width=0.4\textwidth]{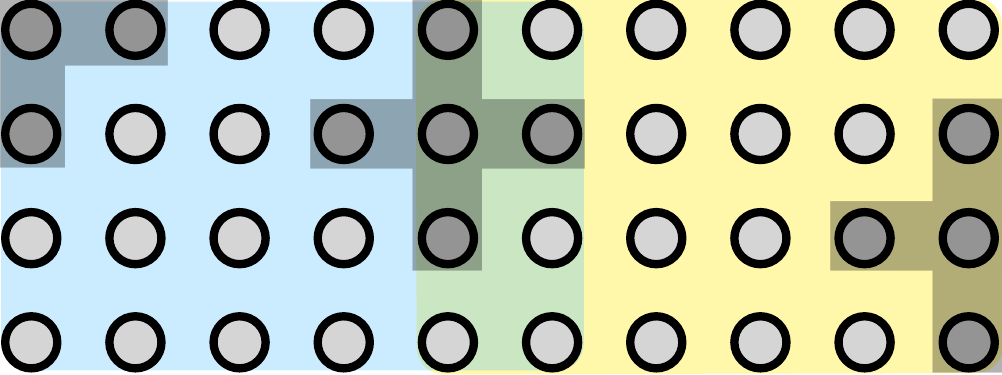}
\caption{Division of the square lattice into $ABC$ for the Hamiltonian $H=\sum_j h_j$ in Eq.~\eqref{eq:GlauberH}. Each term $h_j$, indicated in dark gray, acts on a site $j$ as well as its nearest neighbors. The subsystem $B$ consists of two columns, $B_A$ and $B_C$, so that $A$ and $C$ do not interact directly.}
\label{fig:Glauber}
\end{figure}

To construct our upper bound on the entanglement of $\ket{\psi^H_0}$ let us choose $B$ to consist of two adjacent columns of qubits aligned with the lattice directions, as illustrated in Fig.~\ref{fig:Glauber}. We denote the column closest to $A$ ($C$) by $B_A$ ($B_C$). Two columns are required since we must ensure that there is no interaction between $A$ and $C$. In this notation we have
\begin{align}
	H_{AB} = \sum_{j \in A \cup B_A} h_j, \quad H_{BC} = \sum_{j \in C \cup B_C} h_j.
\end{align}
Crucially, $H_{AB}$ does not involve terms $h_j$ that are `centered' on the qubits $j$ in column $B_C$. The qubits in that column nevertheless appear in the Hilbert space $\mathcal{H}_{AB}$ on which $H_{AB}$ acts. It is straightforward to verify that $[Z_k,H_{AB}]=0$ for $k \in B_C$, and that all of the $2^{L^{d-1}}$ states
\begin{align}
	\ket{\psi^{H_{AB}}_{0,\{z_j \in B_C\}}} \propto \sum_{ \{ z \in A \cup B_A\}} e^{(\beta/4)\sum_j \sum_{k | j} z_j z_k} \ket{\{ z\}},\notag
\end{align}
here indexed by the $L^{d-1}$ bits $z_j$ on sites $j \in B_C$, are ground states of $H_{AB}$. The large ground state degeneracies $D_{AB} = D_{BC} = 2^{L^{d-1}}$ that arise from the missing terms $h_j$ with $j \in B$ give rise to an upper bound on the ground state entanglement entropy of 
\begin{align}
	S_{A,n}(\psi^H_0) \leq L^{d-1} \ln 2,
\end{align}
for all $n \geq 0$. This upper bound scales with the area of the entanglement cut, and matches the maximum possible entanglement for PEPS with bond dimension two, i.e. the bound is saturated by the ground state of $H$. 

Beyond this specific example, the point is that the FF Hamiltonians $H_{AB}$ and $H_{BC}$ of overlapping subsystems will  generally have missing terms near the entanglement cut. These missing terms naturally give rise to degeneracies $D_{AB}$ and $D_{BC}$ that are only exponential in surface areas, and hence lead to area-law upper bounds on ground-state entanglement.

Let us reiterate that the above bounds depend only on ground-state degeneracies of different subsystem Hamiltonians, and not on spectral gaps. To see why such a result is reasonable in a FF system, recall that the ground-state subspace of $H = \sum_j h_j$ is the same as the ground-state subspace of $H_w \equiv \sum_j w_j h_j$ where the weights $w_j > 0$. The spectral gaps, on the other hand, do depend on $w_j$.

\section{Nonzero energies}\label{sec:generic}

Here we discuss our upper bounds on von Neumann entanglement entropy for systems with geometrically local interactions, without requiring that the Hamiltonians are FF. For reasons that will become clear, our focus will be on states at nonzero energy $E-E^H_0$ above the ground state. In Sec.~\ref{sec:technical} we describe the structure of the bounds, and in Sec.~\ref{sec:conventional} we discuss the thermodynamic properties of generic many-body quantum systems. Restricting to systems whose heat capacities are self-averaging, in Sec.~\ref{sec:high} we first construct upper bounds on the entanglement of states with extensive energies. In Sec.~\ref{sec:low} we discuss relations between low-energy entanglement and the way that the specific heat vanishes as the temperature goes to zero.

\subsection{Energetics}\label{sec:technical}

In Sec.~\ref{sec:bounds} we showed how the entanglement in pure states $\ket{\psi}$ with $\braket{\psi|H|\psi}=E$ can be upper bounded by considering a fictitious thermodynamics problem, involving two systems having Hamiltonians $H_{AB}$ and $H_{BC}$. Recall that $H = H_{AB}+H_{BC}$, where $H_{AB}$ and $H_{BC}$ do not commute in general since both act on subsystem $B$ [see Fig.~\ref{fig:mappingfig1}]. In the thermodynamics problem we effectively double the $B$ degrees of freedom, so that $H_{AB}$ acts on $A$ and one of the copies of $B$, while $H_{BC}$ acts on $C$ and the other copy. Here we discuss technical aspects of the fictitious thermodynamics problem and the structure of its solutions.

Having fixed $E$, the first step is to determine the temperature $T^{*}$ of the $H_{AB}$ and $H_{BC}$ systems such that the sum of their energies is $E$. It is convenient to define
\begin{align}
	\epsilon(E) \equiv E - E^{H_{AB}}_0 - E^{H_{BC}}_0.
\end{align}
In terms of the specific heat capacities $c^{H_{AB}}(T)$ and $c^{H_{BC}}(T)$ of these systems we have
\begin{align}
	\epsilon(E)  = \int_0^{T^*} dT \Big[ V_{AB} c^{H_{AB}}(T) + V_{BC} c^{H_{BC}}(T)\Big]. \label{eq:thermalenergy}
\end{align}
The temperature $T^*$ is determined by the condition that the `thermal energy' accounts for the difference $\epsilon(E)$. 

In systems with non-negative heat capacities, for there to be a solution to this equation we must have $\epsilon(E) \geq 0$. To see that this is the case, consider $E=E^H_0$, where $\epsilon(E)$ takes its minimum value. By definition, the expectation value of $H=H_{AB}+H_{BC}$ in the ground state of $H$ is 
\begin{align}
	E^H_0 = \braket{\psi^H_0|H_{AB}|\psi^H_0}+ \braket{\psi^H_0|H_{BC}|\psi^H_0},
\end{align}
while $\braket{\psi^H_0|H_{AB}|\psi^H_0} \geq E^{H_{AB}}_0$, and similar for $H_{BC}$. Therefore $\epsilon(E) \geq 0$ for all $E \geq E^H_0$. All valid constraints on the expectation value $E$ are associated values of $\epsilon(E)$ for which there exist solutions to Eq.~\eqref{eq:thermalenergy}.

In FF systems we have seen that $\epsilon(E^H_0)=0$, but more generally the `frustration' between $H_{AB}$ and $H_{BC}$ leads to $\epsilon(E^H_0)>0$, and hence to a nonzero effective temperature $T^*$ in the upper bounds \eqref{eq:psibound1} and \eqref{eq:psibound2}. The central quantity appearing in those bounds is
\begin{align}
	S^{H_{AB}}(T^*) + S^{H_{BC}}(T^*)\label{eq:psiboundc} = S^{H_{AB}}(0)+S^{H_{BC}}(0)&\\
    +\int_0^{T^*} \frac{dT}{T}\Big[ V_{AB} c^{H_{AB}}(T) + V_{BC} c^{H_{BC}}(T)\Big]&,\notag
\end{align}	
and it is valuable to understand the behavior of this object as $E \to E^H_0$. First, in systems with bounded local Hilbert space dimensions, it is straightforward to prove that $\epsilon(E^H_0) = O(L^{d-1})$. This follows from the facts that $||H_{AB}-H_A||_{\infty}$, $||H_{BC}-H_C||_{\infty}$ and $||H_B||_{\infty}$ are $O(L^{d-1})$, where we define $H_A$ ($H_C$) as the sum of terms acting only on subregion $A$ ($C$). To derive this bound we write
\begin{align}
	\epsilon(E^H_0) &= E^H_0 - E^{H_A + H_C}_0\notag \\
	&+ E^{H_A}_0 - E^{H_{AB}}_0 \\
	&+ E^{H_C}_0 - E^{H_{BC}}_0, \notag
\end{align}
where we have used the fact $H_A$ and $H_C$ commute, and then we use the triangle inequality to upper bound $\epsilon(E^H_0)$ by the sum of moduli of the three lines on the right-hand side. Finally we apply Weyl's inequality to upper bound each term by a spectral norm. For example, $|E^{H_A}_0 - E^{H_{AB}}_0| \leq ||H_{AB}-H_A||_{\infty}$. This leads to
\begin{align}
	\epsilon(E^H_0) = O(L^{d-1}). \label{eq:epsilonbound}
\end{align}
Although we have proved this upper bound only for systems with bounded local Hilbert space dimensions, we expect that it also applies to many bosonic systems. Note that \eqref{eq:epsilonbound} applies regardless of thermodynamic properties, and depends only on the geometric locality of interactions.  

The smallest upper bounds on low-energy entanglement that we can obtain using our framework arise when $T^*$ takes its lowest value. The discussion above indicates that for non-FF systems the scaling of this minimal $T^*$ with $L$ is fixed by the condition that $\epsilon(E) \sim L^{d-1}$ in Eq.~\eqref{eq:thermalenergy}. We then obtain our upper bounds by inserting this $T^*$ into \eqref{eq:psibound1} and \eqref{eq:psibound2}. For FF systems the minimum $T^*=0$ as discussed in Sec.~\ref{sec:frustrationfree}.

\subsection{Conventional thermodynamics}\label{sec:conventional}

The system of equations~\eqref{eq:thermalenergy} and \eqref{eq:psiboundc} simplifies significantly in systems that have conventional thermodynamic properties, in the sense that the heat capacities of $H_{AB}$ and $H_{BC}$ are approximately equal to their thermodynamic limits $c(T)$ when $A$ and $C$ are large, and when the effective temperature $T^*$ is much greater than finite-size energy splittings in the spectrum. Then, given a value of $E$, we will be able to convert this into a value of $T^*$ and an upper bound on entanglement. 

Care is required because we are interested in the scaling of entropies with the system linear extent $L$, and also with temperatures that can \emph{decrease} with $L$. As a first step, let us establish the temperature regimes that we will investigate. Since we expect $\epsilon(E^H_0)=O(L^{d-1})$ in general, we will mostly restrict ourselves to energies $\Omega(L^{d-1})$. We can convert this constraint on the energy into a constraint on the temperature using the fact that the thermal entropy should not diverge in the limit of zero temperature. In terms of the heat capacity, this property implies $c(T) \to 0$ for $T \to 0$, and for energy \emph{densities} $\Omega(L^{-1})$ we then have $T = \Omega(L^{-1})$. 

The finite-size gaps in a many-body spectrum are typically largest near its edges, and there they scale as $\sim L^{-z}$, where the constant $z \geq 1$ in a locally interacting system. Since the temperatures $T=\Omega(L^{-1})$ that we consider can be chosen to be much greater than these finite-size gaps (i.e. by any constant factor), we do not expect the discreteness of the spectrum to lead to drastic differences between, e.g., $c^{H_{AB}}(T)$ and $c(T)$. However, there are boundary effects even where finite-size gaps can be neglected. Because the volume of the boundary region scales as $\sim L^{d-1}$, one generally has
\begin{align}
	c^H(T) = c(T) + \frac{1}{L}b(T) + \cdots, \label{eq:boundaryterm}
\end{align}
and similar for $c^{H_{AB}}(T)$ and $c^{H_{BC}}(T)$. We expect that the boundary contribution $b(T)$ becomes independent of $L$ in large systems. In systems with edge modes, $b(T)$ can have a very different temperature dependence to $c(T)$, but must still vanish as $T \to 0$, since otherwise the boundary would give a divergent contribution to the thermal entropy in finite systems. Due to this constraint, such boundary contributions will be subleading in the examples that we discuss below. 

To summarize, in non-FF systems where $\epsilon(E^H_0)=O(L^{d-1})$ we expect that $T^*$ is sufficiently large that the specific heat capacities are well-approximated by their large $L$ limits $c(T)$. Where we return to FF systems below we will be able to access arbitrarily small values of $T^*$, and for simplicity we will simply neglect finite-size corrections to $c^{H_{AB}}(T)$ and $c^{H_{BC}}(T)$. In both classes of systems we will also assume from here on that the zero-temperature thermal entropies $S^{H_{AB}}(0)$ and $S^{H_{BC}}(0)$ are no larger than $O(L^{d-1})$, although in physical systems that are not FF one generally expects these entropies to only be of order unity.  

\subsection{Finite energy densities}\label{sec:high}

Having discussed technical features of the fictitious thermodynamics problem, we now briefly discuss the case of finite energy densities, $E-E^H_0 \sim L^{d}$. From Eq.~\eqref{eq:epsilonbound} we also have $\epsilon(E) \sim L^d$. For $A$ that makes up half of the overall system, i.e. $V_A = \frac{1}{2}L^d$, $V_B = O(L^{d-1})$, $V_C = L^d-V_A-V_B$, we then have
\begin{align}
	\max^{\psi}_{\braket{H}=E} S_A(\psi) \leq \frac{1}{2}S^H(E) + O(L^{d-1}).\label{eq:extensiveE}
\end{align}
Up to subleading corrections, the maximum possible entanglement entropy of a state with energy $E$ is given by the thermodynamic entropy associated with that energy $E$. 

It has previously been argued that random superpositions of eigenstates of $H$ with energy expectation $E$, as well as the eigenstates themselves in chaotic systems, have half-system entanglement entropies given by $\frac{1}{2}S^H(E)$ up to corrections that are subleading at large $L$ \cite{murthy2019structure,vidmar2017entanglement}. This is one way to see that our bound in \eqref{eq:extensiveE} is optimal up to subleading additive corrections. Notably, in \eqref{eq:extensiveE}, we have shown that \emph{all} quantum states with $\braket{\psi|H|\psi}=E$, including those that have support on eigenstates of $H$ throughout the spectrum, have half-system entanglement entropies upper bounded by half the thermal entropy.

\subsection{Low energies}\label{sec:low}

\begin{figure}
\includegraphics[width=0.47\textwidth]{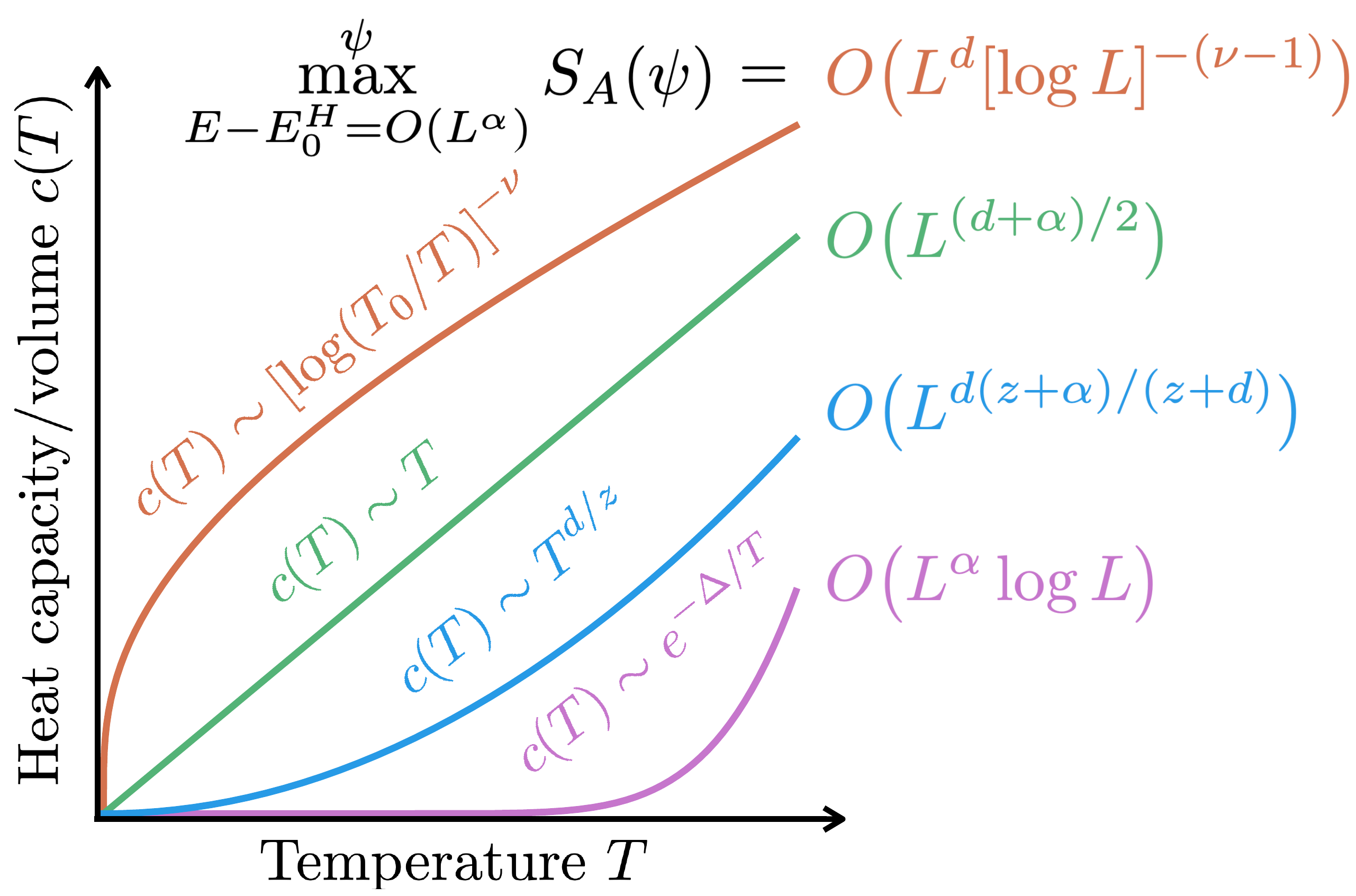}
\caption{Relation between low-temperature specific heat capacity $c(T)$ and the upper bounds on half-system entanglement entropy. Here we show upper bounds for states with energy $E-E^H_0=O(L^{\alpha})$ above the ground state, with $L$ the system linear extent. The scaling of the upper bounds on the right corresponds to an exponent $\alpha$ in the range $d-1 \leq \alpha < d$. The larger the specific heat for $T \to 0$, the faster the growth of the upper bound with $L$. The behaviors displayed are known to arise for (red) certain random quantum critical points, (green) Luttinger and Fermi liquids, (blue) conventional quantum critical points, with $z$ the dynamic critical exponent, and (purple) gapped systems.}
\label{fig:heatcapacitysummary}
\end{figure}

Throughout this section we characterize entanglement in systems at subextensive energies, where behavior is controlled by the temperature dependence of the specific heat capacity in the limit of vanishing temperatures. We will derive upper bounds on entanglement for energies scaling as $E-E^H_0=O(L^{\alpha})$, where $d-1 \leq \alpha < d$ in non-FF systems. Within our framework, and in non-FF systems, bound derived for $\alpha < d-1$ (including those on ground-state entanglement) are no tighter than for $\alpha=d-1$; this follows from Sec.~\ref{sec:technical}. In FF systems, on the other hand, we can extend the analysis down to $E-E^H_0=0$. First in Sec.~\ref{sec:gapped} we discuss gapped systems, and following this in Sec.~\ref{sec:gapless} we study gapless systems whose specific heat follows a power law at low temperatures. Finally in Sec.~\ref{sec:logarithmic} we discuss a class of disordered systems whose specific heat has an inverse-logarithmic dependence on temperature. The results of this section are summarized in Fig.~\ref{fig:heatcapacitysummary}.

\subsubsection{Essential singularity in the heat capacity}\label{sec:gapped}
In gapped quantum phases of matter, in the limit of large $L$ there is by definition a finite separation $\Delta$ between the energy of the ground state subspace and the lowest excited states. Gapped phases arise in wide varities of physical settings, and are characterized by the exponential decay of correlations between local observables \cite{hastings2004lieb}. 

Our model for a gapped phase is a specific heat capacity of the form
\begin{align}
	c(T) = c_0 \Big(\frac{\Delta}{T}\Big)^a e^{-\Delta/T}, \label{eq:specificheatgapped}
\end{align}
which in practice is appropriate for $T \ll \Delta$. As a simple example, in a system whose lowest excited states can be described in terms of weakly interacting quasiparticles, having quadratic dispersion $\omega(\bm{k}) \simeq {\Delta + (2m)^{-1} k^2}$ at small $k=|\bm{k}|$, the exponent $a = 2-d/2$; here $\bm{k}$ is the $d$-component wavevector and $m$ is the effective mass of the quasiparticles. In this case the finite-size gaps at energies just above $\Delta$ scale as $m^{-1} L^{-2}$, so Eq.~\eqref{eq:specificheatgapped} is appropriate for $m^{-1}L^{-2} \ll T \ll \Delta$, as discussed in Sec.~\ref{sec:conventional}.

Now consider a total system energy $E= E^H_0+ W L^{\alpha}$ with respect to a spatially local Hamiltonian $H$ having gap $\Delta$; here $W$ is a constant with dimensions of energy, and the exponent $d-1 \leq \alpha < d$. We choose $V_A = \frac{1}{2}L^d$ and $V_B = O(L^{d-1})$ so that the heat capacities of the $H_{AB}$ and $H_{BC}$ subsystems differ only due to the boundary degrees of freedom. The effective temperature is then given by
\begin{align}
	\frac{1}{2} W L^{\alpha} + \frac{1}{2}\epsilon(E^H_0) + \cdots = \frac{1}{2}L^d \int_0^{T^*} dT c(T) \label{eq:energygapped}
\end{align}
where the ellipsis denotes a boundary correction, which is subleading at large $L$. Here $c(T)$ is given by the low-temperature form in Eq.~\eqref{eq:specificheatgapped} since for any $\alpha < d$ the effective temperature $T^* \ll \Delta$ at large $L$. Recall that for $\alpha < d-1$ we should expect the left-hand side of Eq.~\eqref{eq:energygapped} to be dominated by $\epsilon(E^H_0)=O(L^{d-1})$, as discussed in Sec.~\ref{sec:technical}. Given $T^*$, the thermal entropies on the right-hand side of Eq.~\eqref{eq:psibound1} are
\begin{align}
	S^{H_{AB}}( T^*) = \frac{1}{2}L^d \int_0^{T^*} dT \frac{c(T)}{T} + O(L^{d-1}), \label{eq:thermalentropygapped}
\end{align}
and similar for $S^{H_{BC}}( T^*)$. The $O(L^{d-1})$ correction accounts for the zero-temperature thermal entropy $S^{H_{AB}}(0)$, which we have assumed to grow with $L$ no faster than $\sim L^{d-1}$, as well as for corrections arising from the fact that $V_A$ and $V_C$ may deviate from $\frac{1}{2}L^d$ by $O(L^{d-1})$. Due to the essential singularity in Eq.~\eqref{eq:specificheatgapped} at small $T$, and the fact that here $T^* \ll \Delta$ at large $L$, the integrals in Eqs.~\eqref{eq:energygapped} and \eqref{eq:thermalentropygapped} are dominated by their upper cutoffs. The first term on the right-hand side of Eq.~\eqref{eq:specificheatgapped} is then $r_a \times (E-E^H_0)/T^* + O(L^{d-1})$ where $r_a$ is a dimensionless prefactor that depends on the exponent $a$ in Eq.~\eqref{eq:specificheatgapped}. For example, $r_2=1$. Solving Eq.~\eqref{eq:energygapped} for the effective temperature gives
\begin{align}
	T^* = \frac{\Delta}{(d-\alpha)\ln L} + \cdots,
\end{align}
where the ellipsis denotes contributes that are subleading at large $L$. Inserting this into the above relation between thermal entropy and the energy $E - E^H_0 = W L^{\alpha}$ we find
\begin{align}
	S_A(\psi) \leq \frac{r_a}{2}(d-\alpha)\times \frac{W}{\Delta} L^{\alpha} \ln L + O(L^{d-1}). \label{eq:gappedbound}
\end{align}
Recall that this inequality holds for $d-1 \leq \alpha < d$ in non-FF systems. For FF systems, and neglecting differences between e.g. $c^{H_{AB}}(T)$ and $c(T)$, we can reduce $E$ to $E^H_0$, in which case the first term in \eqref{eq:gappedbound} vanishes. The connection to thermodynamics then shows that the upper bound can be split into a contribution from the ground state, here assumed to be $O(L^{d-1})$, and a contribution from excitations.  

Above we have shown that if a system has a heat capacity of the form in Eq.~\eqref{eq:specificheatgapped}, then the half-system entanglement entropy of a pure state with energy $O(L^{\alpha})$ above the ground state is itself $O(L^{\alpha} \ln L)$ for $d-1 \leq \alpha < d$. For $\alpha=d$ we instead recover volume-law entanglement $O(L^d)$. For $\alpha=d-1$, where we find our most restrictive upper bound applicable to generic (non-FF) systems, the result is
\begin{align}
	\max^{\psi}_{\braket{H}=E^H_0+O(L^{d-1})} S_A(\psi) = O\big( L^{d-1} \ln L\big), \label{eq:gappedgroundstate}
\end{align}
which is the area law up to a multiplicative logarithmic correction. This is an upper bound on the ground state entanglement entropies of gapped ground states in general spatial dimensions, but we note that it also applies to all energies $E$ scaling with the boundary of a subregion. 

Previous work \cite{masanes2009area} proved an upper bound of the form in Eq.~\eqref{eq:gappedgroundstate} for gapped ground states using very different methods and assumptions. Later, \cite{brandao2015entanglement} arrived at the same scaling through a closely related mapping to thermodynamics. We now show that the upper bound in Eq.~\eqref{eq:gappedbound}, including the prefactor, is saturated by certain random states in a system with $a=2$ in Eq.~\eqref{eq:specificheatgapped}.

We consider a system of decoupled qubits ${H = -\frac{1}{2}\Delta \sum_j Z_j}$, and study the entanglement of random superpositions of states in sectors with fixed $\sum_j Z_j$, i.e. states with zero energy variance and $E-E^H_0=k \Delta$ with $k \sim L^{\alpha}$ a positive integer with $d-1 \leq \alpha < d$. In a system of $L^d$ spins the dimension of the space with $E-E^H_0=k \Delta$ is $L^d \choose k$. Setting $k = \lceil (W/\Delta) L^{\alpha} \rceil$ the average von Neumann entropy of a random state with $\sum_j Z_j = k-L$ is then \cite{vidmar2017entanglement}
\begin{align}
	\bar{S}_A = \frac{1}{2}\frac{W}{\Delta} (d-\alpha) L^{\alpha} \ln L + \cdots \label{eq:gappedrandomstate}
\end{align}
at large $L$. To compare this with our upper bound \eqref{eq:gappedbound}, note that for decoupled qubits $c(T) = [\Delta/T]^2 e^{-\Delta/T} + \cdots$ where the ellipsis denotes contributions that are subleading as $T$ goes to zero. Since $r_{2}=1$, the leading terms in \eqref{eq:gappedbound} and \eqref{eq:gappedrandomstate} match. Our general result in \eqref{eq:gappedrandomstate} is therefore optimal up to corrections that are subleading at large $L$. 

Although we have derived \eqref{eq:gappedgroundstate} from purely thermodynamic considerations, the additional assumption that low-energy states can be described in terms of quasiparticles allows for a physical interpretation of the scaling with $L$. First, in the simplest case of $d=1$, for $\alpha=0$ and hence an amount of energy $W$, it is possible to create $\sim W/\Delta$ quasiparticles. These excitations can be in any of $\sim L$ locations. The effective Hilbert space dimension at $E \leq W$ is therefore $\sim L^{W/\Delta}$, and a random state in a space of this dimension has entanglement entropy $S_A \sim [W/\Delta] \ln L$. More generally, in $d$ dimensions and with energy $E=W L^{\alpha}$ and $\alpha < d$, there are up to $\sim [W/\Delta] L^{\alpha}$ quasiparticles, so the effective Hilbert space dimension is $\sim (L^d)^{[W/\Delta] L^{\alpha}}$. Note that this approximation is only appropriate for $\alpha$ strictly smaller than $d$, since then quasiparticles are `dilute' in space.  

In this section, as a first application of our framework to generic systems, we have derived upper bounds on entanglement for gapped systems. An essential singularity in the low-temperature specific heat $c(T) \sim e^{-\Delta/T}$ leads to an upper bound on entanglement that is proportional to (i) the energy above the ground state and (ii) a logarithmic term $\sim \ln L$ which can be interpreted as the configurational entropy of a quasiparticle excitation.

\subsubsection{Power-law heat capacity}\label{sec:gapless}

Quantum systems with gapless branches of excitations often have low-temperature heat capacities that grow with a power of temperature. Key examples are metals whose lowest-energy excitations are associated with transitions across a Fermi surface, quantum critical systems \cite{sachdev2011quantum}, and certain quantum spin liquids \cite{motrunich2005variational}. Power-law heat capacities can also arise in disordered systems due to Griffiths effects \cite{thill1995equilibrium,guo1996quantum,rieger1996griffiths} or tunneling between different low-energy states \cite{phillips1972tunneling,anderson1972anomalous}. The presence of low-energy excitations will allow for far more entanglement at subextensive energies than in gapped systems.

Here we study entanglement in systems with heat capacities of the form
\begin{align}
	c(T) = c_0 T^{\gamma} + \cdots \label{eq:gaplessheatcapacity}
\end{align}
where the ellipsis denotes contributions that are subleading when the energy density is $\Omega(L^{-1})$. In lattice models, heat capacities of the form \eqref{eq:gaplessheatcapacity} generally arise at temperatures that are small compared with microscopic energy scales, while at high temperatures the specific heat capacity approaches a constant. 

In metals one finds a linear-in-$T$ specific heat, i.e. with $\gamma=1$. The exponent $\gamma=1$ is also realized in wide varieties of glassy systems \cite{phillips1972tunneling,anderson1972anomalous}. In quantum critical systems, if the lowest excitations are weakly interacting and have dispersion $\omega \sim k^z$, where $z$ is the dynamic critical exponent, then the exponent $\gamma = d/z$, where $d$ is the spatial dimension. We note that our results only depend on the scaling of the heat capacity with temperature, and not on the existence of excitations having well-defined wavenumber. For example, our results apply also to certain disordered systems, as well as to systems without well-defined quasiparticle excitations, provided the low-temperature heat capacities follow a power law. 

Given an energy $E$ and hence $\epsilon = \epsilon(E)$, from Eq.~\eqref{eq:thermalenergy} the effective temperature is given by
\begin{align}
	T^* = \Bigg[\frac{2(\gamma+1)\epsilon}{c_0 L^d}\Bigg]^{1/(\gamma+1)}.
\end{align}
We then find an upper bound on the entanglement entropy for states $\ket{\psi}$ with $\braket{\psi|H|\psi}=E$ of the form
\begin{align}
	S_A(\psi) \leq r'_{\gamma} \Big[L^d \epsilon(E)^{\gamma} \Big]^{1/(\gamma+1)} + O(L^{d-1}), \label{eq:upperboundgapless}
\end{align}
where the coefficient $r'_{\gamma} = \gamma^{-1} [2^{-1} (\gamma+1)^{\gamma} c_0]^{1/(\gamma+1)}$ increases with the number of low-energy excitations. Since $\epsilon(E^H_0) \sim L^{d-1}$ in non-FF systems, the $O(L^{d-1})$ term is subleading. In FF systems we are allowed to reduce $\epsilon(E)$ to be arbitrarily small, so that the upper bound is dominated by the ground-state contribution $O(L^{d-1})$.

Setting $E-E^H_0 = O(L^{\alpha})$ with $d-1 \leq \alpha \leq d$ in \eqref{eq:upperboundgapless} we find that the entanglement entropy is upper bounded as
\begin{align}
	\max^{\psi}_{\braket{H}=E^H_0+O(L^{\alpha})} S_A(\psi) = O\Big( L^{(d+\alpha \gamma)/(1+\gamma)}\Big).\label{eq:gaplessbound}
\end{align}
Recall that for any $\alpha < d$ the effective temperature $T^*$ decreases with $L$, meaning that the fictitious thermodynamic problem is in the low-temperature regime where Eq.~\eqref{eq:gaplessheatcapacity} is appropriate. 

In generic systems the bound \eqref{eq:gaplessbound} is tightest for $\alpha=d-1$, where the right-hand side is $O\big(L^{d - \gamma/(1+\gamma)}\big)$. We focus on $\alpha=d-1$ for the remainder of this subsection. In metallic phases the existence of a Fermi surface generically leads to $\gamma=1$, so that
\begin{align}
	\max^{\psi}_{\braket{H}=E^H_0+O(L^{d-1})} S_A(\psi) = O\Big( L^{d-1} \times L^{1/2}\Big). \label{eq:metal}
\end{align}
At quantum critical points with dynamic critical exponent $z$ one instead finds $\gamma = d/z$, leading to
\begin{align}
	\max^{\psi}_{\braket{H}=E^H_0+O(L^{d-1})} S_A(\psi) = O\Big( L^{d-1} \times L^{z/(z+d)}\Big). \label{eq:critical}
\end{align}
Note that for $d=z=1$ the upper bounds in \eqref{eq:metal} and \eqref{eq:critical} match; this scenario corresponds to a Luttinger liquid, which can be described as either an interacting system of fermions or as a bosonic quantum critical system \cite{haldane1981luttinger}. 

Above we have discussed the maximum possible entanglement entropies of quantum states with fixed energy. Work on quantum thermalization suggests that random quantum states at finite energy densities have subsystem entanglement entropies close to the subsystem thermal entropies (at the corresponding energies). Moreover, in generic interacting systems, eigenstates resemble random states (this is the content of the ETH \cite{deutsch1991quantum,srednicki1994chaos}). We now show numerically that even in non-interacting systems of fermions, where eigenstates are Slater determinants, there exist families of low-energy eigenstates whose entanglement entropies follow the scaling predicted in our upper bound \eqref{eq:metal}, although with a modified prefactor. These eigenstates are associated with large numbers of low-energy excitations across the Fermi surface. 

\begin{figure}
\includegraphics[width=0.47\textwidth]{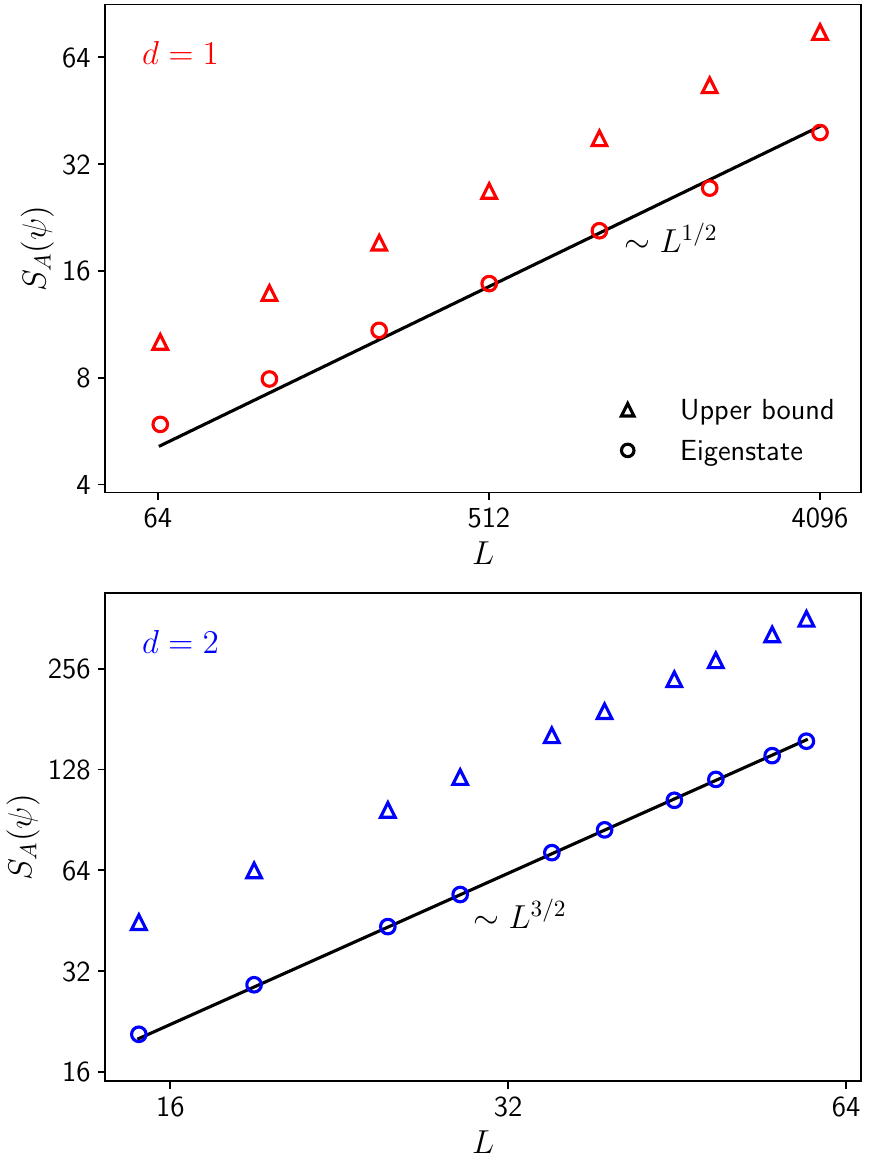}
\caption{Comparing upper bounds on entanglement (triangles) with eigenstate entanglement (circles) in systems of free fermions in spatial dimension $d=1$ (above) and $d=2$ (below). The geometry is as shown in Fig.~\ref{fig:mappingfig1}. Solid lines show the scaling predicted $\sim L^{d-1/2}$ in \eqref{eq:metal}, and here we fit the prefactor and vertical offset to the numerical results on eigenstate entanglement. Eigenstates $\ket{\psi}$ have definite parity under reflection about $B$, so $S_A(\psi)=S_C(\psi)$ and the upper bound displayed is half the right-hand side of \eqref{eq:psibound2}. We evaluate this upper bound via exact diagonalization of $H_{AB}$ (and $H_{BC}$). For $d=1$ and $d=2$ we randomly sample $10^3$ eigenstates with $E-E^H_0 \leq 5$ and $E-E^H_0 \leq L$, respectively, and show the maximum $S_A(\psi)$ over these samples.  }
\label{fig:metalnumerics}
\end{figure}

For concreteness we study the standard tight-binding model 
\begin{align}
	H = -\sum_{\langle j,k\rangle} c_j^{\dag} c_k + \text{H.c.}, \label{eq:Hfermions}
\end{align}
where $c_j$ and $c_j^{\dag}$ are fermionic annihilation and creation operators, respectively, at site $j$ of the lattice. The sum is over nearest neighbors $\langle j,k\rangle$. Our results in $d=1$ correspond to a one-dimensional chain with open boundary conditions and $L$ sites, while in $d=2$ we study an isotropic square lattice with open boundary conditions and linear dimension $L$ (corresponding to $L^2$ sites in total). We restrict $L$ to be odd, so the subsystem $A$ ($C$) then corresponds to the first (last) $\lfloor L/2\rfloor$ sites for $d=1$ and the first (last) $\lfloor L/2\rfloor$ columns for $d=2$. Subsystem $B$ consists of the site (in $d=1$) or column (in $d=2$) in the center of the system, which buffers $A$ and $C$. This is the geometry shown in Fig.~\ref{fig:mappingfig1}. We then have, for example,
\begin{align}
	H_{AB} = -\sum_{\langle j,k\rangle \in AB} \Big( c_j^{\dag} c_k + c_k^{\dag} c_j\Big), \label{eq:HABfermions}
\end{align}
and similar for $H_{BC}$. 

For various $L$ we randomly sample $10^3$ many-body eigenstates constrained to have ${E-E^H_0 \leq 5}$ for $d=1$, or ${E-E^H_0 \leq L^{d-1}}$ for $d=2$, and we compute the maximum over samples of their half-system entanglement entropies. Note that $E^H_0$ is the ground state energy of the full Hamiltonian $H$, rather than of its restriction to a specific particle number sector, and that the eigenstates that we sample are similarly not constrained to have a specific particle number. By varying $L$, in Fig.~\ref{fig:metalnumerics}, we show that for the most highly entangled eigenstates the entanglement entropy $S_A(\psi)$ grows with $L$ with (approximately) the power appearing in the upper bound $S_A(\psi) = O(L^{d-1} \times L^{1/2})$ in \eqref{eq:metal}. 

Because the systems that we have studied have the geometry in Fig.~\ref{fig:mappingfig}, as well as reflection symmetry about $B$, the results of Sec.~\ref{sec:lowerbound} imply that there exist states with $E-E^H_0=O(L^{d-1})$ that have entanglement $S_A(\psi) = O(L^{d-1} \times L^{1/2})$. More precisely, there exist pure states whose entanglement entropies $S_A(\psi)$ match the upper bound up to an additive correction that is small at large $L$. The families of eigenstates studied in Fig.~\ref{fig:metalnumerics} have $S_A(\psi) = O(L^{d-1} \times L^{1/2})$ but the prefactor does not match the upper bound, i.e. the entanglement entropies of these eigenstates are lower than the upper bound by a multiplicative rather than just an additive correction. This is unsurprising because the eigenstates of the Hamiltonian studied here are simple Slater determinants, while generic pure states are superpositions of such Slater determinants, and these superpositions can be more highly entangled.

It is useful to contrast the scaling above with that in glassy systems where $c(T) \sim T$ is expected to arise from isolated two-level systems (TLSs) \cite{phillips1972tunneling,anderson1972anomalous}. These TLSs are typically understood as corresponding to pairs of locally distinct configurations of spins or particles, whose energy differences $\Delta E$ are drawn from a distribution with nonzero weight near $\Delta E = 0$. A phenomenological Hamiltonian is simply $H = \sum_j (\Delta E)_j Z_j$, where here each $Z_j$ operator is associated with a distinct TLS (and so is extended over a finite region of space). If $\Delta E_j \gg T$ the TLS $j$ is not thermally active at temperature $T$, whereas if $\Delta E_j \ll T$ it contributes $\sim (\Delta E_j/T)^2$ to the heat capacity. Averaging $\Delta E_j$ over space one finds $c(T) \sim T$. In this picture the eigenstates are products over TLSs, and only those local TLSs whose configurations cross the entanglement cut will contribute to entanglement, meaning that $S_A \sim L^{d-1}$. Random states at a given energy $E-E^H_0 \sim L^{\alpha}$ will on the other hand have $S_A \sim \sqrt{L^d E}$. In Fig.~\ref{fig:metalnumerics} we saw that in free fermion systems, the scaling $S_A \sim \sqrt{L^d E}$ is instead obeyed by the eigenstates because low-energy excitations are spatially delocalized.

\subsubsection{Inverse-logarithmic heat capacity}\label{sec:logarithmic}

Spatial disorder is naturally associated with the emergence of low-energy quantum degrees of freedom, and these can lead to exotic thermodynamic properties. For example, in spin chains described by infinite-randomness fixed points \cite{fisher1994random}, one finds \cite{fisher1995critical}
\begin{align}
	c(T) = \frac{c_0}{\ln^{\nu}(T_0/T)} + \ldots
\end{align}
where, as above, the ellipsis denotes contributions that are subleading for energy densities that are $\Omega(L^{-1})$. Although the above behavior has so far been identified in $d=1$, in this section we will study the consequences of this logarithmic divergence for general $d$. We will also restrict to $E-E^H_0 = O(L^{\alpha})$ with $\alpha$ strictly smaller than $d$. 

From Eq.~\eqref{eq:thermalenergy} we find that the effective temperature $T^*$ is given by the condition
\begin{align}
	\epsilon(E) = c_0 L^d T_0 \int_{\ln(T_0/T_*)}^{\infty} du\, u^{-\nu} e^{-u}.
\end{align}
leading to $T_* \sim L^{-(d-\alpha)}$ up to logarithmic multiplicative corrections. This implies a thermal entropy per unit volume of
\begin{align}
	\int_{\ln(T_0/T_*)}^{\infty} dT c(T) = \frac{1}{2}L^d c_0 \big[ (d-\alpha) \ln L \big]^{1-\nu}+\ldots,
\end{align}
at large $L$. The upper bound on the entanglement entropy is then
\begin{align}
	\max^{\psi}_{\braket{H}=E^H_0+O(L^{\alpha})} S_A(\psi) = O\Big(L^d \times \big[\ln L\big]^{-(\nu-1)}\Big), \label{eq:inverselog}
\end{align}	
for $d-1 \leq \alpha < d$. Interestingly, the large density of states at low energies allows for states that have subextensive energies and entanglement entropies that a follow a \emph{volume law} up to a multiplicative logarithmic correction. 

As a concrete example, random transverse-field Ising models in $d=1$ have $\nu=3$ at their critical point \cite{fisher1995critical}, and setting $\alpha=0$ we find an upper bound on the entanglement entropies of systems where the energy above the ground state is of order unity. The upper bound is then
\begin{align}
	\max^{\psi}_{\braket{H}=E^H_0+O(L^{\alpha})} S_A(\psi) \leq \frac{1}{2}c_0 L \big[ \ln L \big]^{-2} + \ldots \label{eq:randomsinglet}
\end{align}
By contrast, low-energy eigenstates of this model are expected to be described qualitatively by random singlet physics; this picture leads to a ground-state entanglement entropy which scales logarithmically with $L$ \cite{refael2004entanglement}. We expect that the upper bound in \eqref{eq:randomsinglet} is saturated (up to corrections that are subleading at large $L$) for superpositions of large numbers of low-energy eigenstates.

\section{Zero-temperature response}\label{sec:response}

The low-energy excitations of quantum condensed matter systems can often be analyzed via zero-temperature response functions. It is therefore important to ask about the complexity of calculating such quantities using classical computers. One way to calculate a zero-temperature response function is to (i) represent the quantum state as a tensor network, (ii) perturb this state with a local observable (or a linear combination of local observables), thereby injecting an amount of energy of order unity, and (iii) evolve the resulting state under the Hamiltonian. The response of various observables can then be computed from a representation of the time-evolved state. 

Using the formalism developed in the previous sections, here we identify situations where at large $L$ the maximum von Neumann entanglement entropy $S_A(\psi)$ at $E-E^H_0 = O(1)$ is dominated by the ground-state contribution. Our rigorous results here apply for FF systems and arbitrary low-temperature heat capacities (corresponding to both gapped and gapless spectra), since in this case our upper bound on entanglement continues to decrease as $E \to E^H_0$ from above. In the following we will make extensive use of the fact that, in the fictitious thermodynamics problem relevant to our upper bound on entanglement, we have captured the energy constraint exactly. 

On decreasing $E$ in a non-FF system the upper bound on entanglement stops decreasing when $E - E^H_0 \sim L^{d-1}$. This is why, for example, our upper bound on ground-state entanglement is $\sim L^{d-1}\ln L$ (see also Refs.~\cite{masanes2009area,brandao2015entanglement}), whereas one expects $\sim L^{d-1}$ in practice. In non-FF systems we are therefore unable to cleanly separate out the ground-state contribution to low energy entanglement. We nevertheless expect that the lessons learned from our analysis of FF systems hold more generally, since they essentially come from counting the number of low-energy excited states. We will also perform numerical calculation to provide indirect support for this claim.

First we discuss gapped FF systems. Fixing $E -E^H_0 = O(1)$, an essential singularity in the heat capacity $c(T) \sim e^{-\Delta/T}$ leads to an effective temperature in the fictitious thermodynamics problem that is logarithmic in system size $T^* \sim \Delta^{-1} \ln L$. The upper bound on entanglement above the ground state for $\braket{\psi|H|\psi}=E$ is then
\begin{align}
	S_A(\psi) - S_A(\psi^H_0) = O\big( \ln L \big), \label{eq:excitationentropy1}
\end{align}
which is subleading relative to the area law $S_A(\psi^H_0) \sim L^{d-1}$ in dimension $d \geq 2$. This comparison between ground-state and excitation contributions to entanglement is more subtle in gapless systems. A power-law specific heat capacity $c(T) \sim T^{\gamma}$ implies an effective temperature $T^*$ scaling with the system linear extent as $T^* \sim L^{-d/(1+\gamma)}$ when $E-E^H_0=O(1)$, so for $\braket{\psi|H|\psi}=E$ we find
\begin{align}
	S_A(\psi) - S_A(\psi^H_0) = O\big( L^{d/(1+\gamma)}\big) \label{eq:excitationentropy2}
\end{align}
and we can view the right-hand side as the contribution from excitations. Even if the ground state is area-law entangled, the excitation contribution in Eq.~\eqref{eq:excitationentropy2} is subleading for $\gamma > (d-1)^{-1}$.

Recall that if the power-law heat capacity arises from bosonic excitations with dispersion $\omega \sim k^z$ in $d$ spatial dimensions, the exponent $\gamma = d/z$. This is important because bosonic systems with gapless excitations and large $z$ are rather uncommon. In a coarse-grained field theory description, achieving $\gamma < (d-1)^{-1}$ in $d = 2$ requires $z > 2$, and therefore the vanishing of lowest-order spatial gradients, while in $d=3$ the requirements are far more extreme: one would need $z > 6$ for the excitation contribution to overwhelm the ground state. 

Quantum criticality with $z > 2$ in $d=2$ is nevertheless expected to arise in metallic systems where low-energy fermionic excitations are coupled to a gapless bosonic field \cite{hertz1976quantum,millis1993effect}. Dispersions $\omega \sim k^z$ with large $z$ are also known to arise in certain multi-layer materials \cite{zhang2010band}. Physically the excitation contribution to entanglement can dominate if, after locally injecting a finite amount of energy, interactions cause the initial excitation to break up into many low-energy degrees of freedom. This can be seen from the fact that the thermal entropy in Eq.~\eqref{eq:excitationentropy2} is dominated by states with many excitations. 

Above, frustration-freeness and the low-temperature behavior of the specific heat capacity were \emph{inputs} to the calculation. Whether these two inputs are consistent with one another is a separate question. Notably, Refs.~\cite{masaoka2025rigorous,masaoka2026frustration,lemm2025critical} showed that gapless FF systems whose spatial correlations decay as power laws must have excitation gaps that decay at least as fast as $L^{-2}$ (up to logarithmic corrections). If the density of states of the corresponding excitations resembles that of quasiparticles, the low-temperature specific heat $c(T) \sim T^{\gamma}$ would have exponent $\gamma \leq d/2$.

Although our upper bounds for $E-E^H_0 =O(1)$ in $d \geq 2$ are rigorous only for FF systems, it is possible that they hold more generally. That said, since we are ultimately interested in the scaling of the various contributions to entanglement with $L$ in more than one spatial dimension, the numerical techniques that we can use to test this idea are unfortunately rather limited. Recall that in Fig.~\ref{fig:metalnumerics} we nevertheless found that certain families of eigenstates of non-interacting fermion systems have half-system entanglement entropies whose scaling with $L$ matches our general upper bound in the regime $E-E^H_0 = O(L^{d-1})$. In particular, in $d=2$ we showed that there are eigenstates with $E \sim L$ and $S_A(\psi) \sim L^{3/2}$.

\begin{figure}
\includegraphics[width=0.47\textwidth]{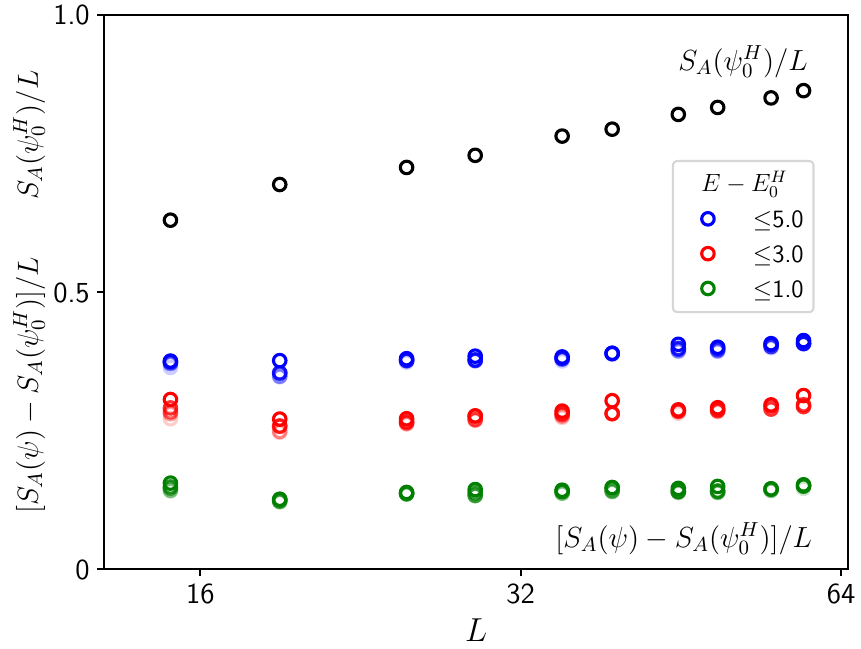}
\caption{Contributions to the half-system entanglement of eigenstates of the non-interacting fermion systems in Eq.~\eqref{eq:Hfermions} in dimension $d=2$. For each $L$ we sample $10^3$ random eigenstates $\ket{\psi}$ having energy within $E-E^H_0$ of the ground state, and for each choice of $E-E^H_0$ (different colors, see legend) we show the five largest excitation contributions $S_A(\psi)-S_A(\psi^H_0)$ to the half-system entanglement $S_A(\psi)$. The scaling of this contribution with $L$ is compared with that of the ground-state entanglement $S_A(\psi^H_0)$ (black) which grows as $\sim L \ln L$. Results are scaled by $1/L$ in order to highlight the different $L$-dependencies of the ground $S_A(\psi^H_0)$ and excitation contributions $S_A(\psi)-S_A(\psi^H_0)$.}
\label{fig:excitationcontribution}
\end{figure}

For the (non-FF) two-dimensional system of fermions in Eq.~\eqref{eq:Hfermions} we now show that the entanglement entropies $S_A(\psi)$ of eigenstates with $E-E^H_0=O(1)$ are dominated by the ground-state contribution. In Fig.~\ref{fig:excitationcontribution} for various $L$ we compute the maximum excitation contribution $S_A(\psi)-S_A(\psi^H_0)$ to entanglement over $10^3$ randomly sampled eigenstates $\ket{\psi}$ whose energies $E$ are below a threshold indicated on the legend, and we compare this maximum with the ground-state entanglement. We also show that the next four largest excitation contributions have essentially the same scaling with $L$. 

Recall that $S_A(\psi^H_0) \sim L \ln L$ while our calculations for FF systems indicate that the excitation contribution $S_A(\psi) - S_A(\psi^H_0)$ in Eq.~\eqref{eq:excitationentropy2} is at most $O(L)$ since here $\gamma=1$. The behavior in Fig.~\ref{fig:excitationcontribution} is consistent with this: the scaled excitation contribution $[S_A(\psi) - S_A(\psi^H_0)]/L$ appears to approach a constant, which is to be contrasted with the clear logarithmic growth of $S_A(\psi^H_0)/L$ with $L$.

This result supports the idea that the excitation contribution to entanglement can generally be constrained by a mapping to a fictitious thermodynamics problem, although we reiterate that we have only proved that this is the case for FF systems. The implication would be that in $d \geq 2$ the finite-energy dynamics of wide varieties of locally interacting many-body quantum systems occurs within a submanifold of Hilbert space in which $S_A(\psi)$ is dominated by the ground-state contribution. Notable exceptions arise when low-temperature heat capacities grow with small powers of the temperature.

When finite-energy many-body states are area-law entangled, and when ground states can be approximately described by tensor networks with finite bond dimension, it is natural to conjecture that finite-energy dynamics can also be efficiently simulated. Indeed, advances in numerical algorithms for real-time evolution have recently allowed for simulations of local quenches in two spatial dimensions~\cite{wu2025accurate,wu2026real}.

\section{Discussion}\label{sec:discussion}

A central challenge in many-body quantum mechanics is to relate phenomena that we observe in experiments to the mathematical structure of quantum states. For example, a simple explanation for the thermalization of isolated quantum systems is that many-body eigenstates resemble random vectors \cite{deutsch1991quantum,srednicki1994chaos}. Scale-invariant correlations at quantum critical points, on the other hand, necessitate entanglement in the corresponding ground states \cite{calabrese2004entanglement}. In this work we have shown how thermodynamic properties, in particular low- and zero-temperature thermal entropies of subsystems, constrain the structure of many-body Hilbert space at low energies. 

Our main result came from relaxing a constrained optimization problem: the maximum entanglement entropy of a quantum state with fixed energy can be upper bounded by a linear combination of entropies of states of overlapping subsystems, where the subsystem states are not necessarily consistent with any state of the entire system. This result applies to Hamiltonians with geometrically local interactions and without any symmetry restrictions. Since entanglement is related to the computational resources required to represent quantum states as tensor networks \cite{giufre2003efficient}, and energy is conserved under Hamiltonian dynamics, our bounds provide insights into the complexity of simulating time evolution on classical computers.

For FF systems we showed that, if the zero-temperature thermal entropy of a subsystem grows with its surface area but not with its volume, then ground states of the full system are area-law entangled. This follows from an upper bound on Schmidt ranks that is expressed in terms of ground-state degeneracies on overlapping subsystems. The fact that the upper bound depends only on ground-state subspaces, and not on spectral gaps, can be understood as a consequence of the fact that in FF systems these subspaces are invariant under the independent rescaling of local Hamiltonian terms by positive numbers. Such rescalings can, however, change spectral gaps. 

As a consequence of our upper bound on the Schmidt rank, the only way for ground-state entanglement to exceed area-law scaling is if the ground state \emph{degeneracies} of subsystems grow faster than exponentially with the subsystem surface area. This is the reason why the ground states studied in e.g. Refs.~\cite{movassagh2016supercritical,zhang2017novel} are highly entangled. However, such large degeneracies are not expected to be generic because they are unstable to perturbations. 

In systems whose heat capacities are self-averaging, and at energies that scale subextensively with system size, the density of states at low energy controls the scaling of our upper bound on entanglement entropy. Moreover, for energies that are $\Omega(L^{d-1})$ above the ground state, we have been able to prove that the upper bounds are optimal up to an area-law correction for a large class of quantum systems. This result is fairly intuitive: the lower the energies of excitations, the greater the number of excitations that can be created in a system with a fixed total energy. As the number of possible configurations of these excitations increases, so does the maximum allowed entanglement entropy of a quantum state. 

Finally we proved that, in FF systems with conventional thermodynamics properties, many-body quantum states having energies of order unity are typically rather weakly entangled. For example, in two spatial dimensions, and for a heat capacity that is linear in temperature, the entanglement entropies of such states can only exceed that of the ground state by an area-law contribution. 

This connection has potentially far-reaching implications for simulations of quantum condensed matter systems. Many quantum phases of matter have geometrically local FF Hamiltonian representatives, such as quantum ferromagnets, valence-bond solids \cite{affleck1988valence}, and the wide variety of topological phases arising in the context of the fractional quantum Hall effect \cite{nielsen2013local}. Since our result bounds the amount of entanglement that can be generated in simulations of zero-temperature response, it constrains the computational resources required for tensor-network representations of quantum states. This is particularly useful if the phase has a FF representative whose low-energy dynamics provide insights into universal properties of excitations. However, this is not always the case: it may be that frustration-freeness leads to a qualitative change in dynamics relative to the generic case (obvious examples are commuting Hamiltonians), or even that a FF representative does not exist. A pertinent question is therefore whether our rigorous upper bounds on entanglement at energies of order unity can be extended, in spatial dimension two and higher, to systems that are not FF.

It is also natural to ask about the implications of our results for the structure of energy eigenstates. While the ETH is typically expected to provide a qualitative description of eigenstates whose energies are extensive, it is unclear whether it is appropriate at subextensive energies. Our work shows that there are pure states that have subextensive energies and half-system entanglement entropies that scale with the thermodynamic entropy, but does not tell us anything about the relations between these pure states and energy eigenstates.

Studies of entanglement dynamics can provide indirect insights into these relations. A key question is whether quenches from weakly entangled low-energy states (in particular, states with subextensive energies) lead to the emergence of maximally entangled states on physical time scales. This is closely related to the question of how spatially dilute quasiparticles scatter off of one another and decay, a process that has been investigated extensively via the Boltzmann equation \cite{bonitz2016quantum}, and more recently via exact calculations of chaos diagnostics in semiclassical systems~\cite{ruidas2026many}. Away from semiclassical limits, numerical calculations based on tensor networks are uniquely suited to this problem because low-energy states are weakly entangled compared with generic states. 

\acknowledgements

The authors are grateful to Yantao Wu, Ehud Altman, Amanda Young, Jerome Dubail, Sarang Gopalakrishnan, Mike Zaletel and Mari-Carmen Ba\~nuls for useful discussions, and to Max McGinley for useful comments on the manuscript. This work was supported by a Brown Investigator Award, a program of the Brown Institute for Basic Sciences at the California Institute of Technology.

%

\end{document}